\documentclass[letterspace]{article}
\pdfoutput=1

\usepackage{jheppub,setspace}
\usepackage{url,comment}
\usepackage{times}
\usepackage{latexsym}
\usepackage{graphicx, graphics, hyperref, amsmath, amssymb, slashed, color, bbm,amsthm, array}
\usepackage[usenames,dvipsnames]{xcolor}
 \usepackage{subfigure}
\usepackage{mdwlist}
\usepackage{multirow}

\numberwithin{equation}{section}

\def \beg#1{\begin{#1}} 
\def \be{\beg{equation}}
\def \bea{\beg{eqnarray}}
\def \eea{\end{eqnarray}}
\def \ee{\end{equation}}

\def\bc{\begin{center}}
\def\ec{\end{center}}

\def\nn{\nonumber}

\title{Dynamics of multi-stable states during ongoing and evoked cortical activity\linebreak
{\normalsize {\it J Neurosci} \normalfont 35(21): 8214-8231, 2015}
}

\author[1]{Luca Mazzucato}
\author[1,2,*]{Alfredo Fontanini}
\author[1,2,*]{and Giancarlo La Camera}

\affiliation[1]{Department of Neurobiology and Behavior and}
\affiliation[2]{Graduate Program in Neuroscience, State University of New York at Stony Brook, Stony Brook, NY 11794}
\affiliation[*]{co-corresponding authors}
  
\bigskip
\abstract{  
Single trial analyses of ensemble activity in alert animals demonstrate that cortical circuits dynamics evolve through temporal sequences of metastable states. Metastability has been studied for its potential role in sensory coding, memory and decision-making. Yet, very little is known about the network mechanisms responsible for its genesis. It is often assumed that the onset of state sequences is triggered by an external stimulus. Here we show that state sequences can be observed also in the absence of overt sensory stimulation. Analysis of multielectrode recordings from the gustatory cortex of alert rats revealed ongoing sequences of states, where single neurons spontaneously attain several firing rates across different states. This single neuron multi-stability represents a challenge to existing spiking network models, where typically each neuron is at most bi-stable. We present a recurrent spiking network model that accounts for both the spontaneous generation of state sequences and the multi-stability in single neuron firing rates. Each state results from the activation of neural clusters with potentiated intra-cluster connections, with the firing rate in each cluster depending on the number of active clusters. Simulations show that the modelÕs ensemble activity hops among the different states, reproducing the ongoing dynamics observed in the data. When probed with external stimuli, the model predicts the quenching of single neuron multi-stability into bi-stability and the reduction of trial-by-trial variability. Both predictions were confirmed in the data. Altogether, these results provide a theoretical framework that captures both ongoing and evoked network dynamics in a single mechanistic model.
\

\bigskip
\noindent Emails: $\textsf{luca dot mazzucato}$, $\textsf{alfredo dot fontanini}$, $\textsf{giancarlo dot lacamera at stonybrook.edu}$
\noindent 
}


\begin{document}

\setcounter{tocdepth}{2}
\maketitle

\section{Introduction}

Sensory networks respond to stimuli with dynamic and time-varying modulations in neural activity. Single trial analyses of sensory responses have shown that cortical networks may quickly transition from one state of coordinated firing to another as a stimulus is being processed or a movement is being performed \cite{Abeles1995a, Seidemann1996,Laurent2001,Baeg2003,Lin2005,Rabinovich2008, PonceAlvarez2012}. Applying a Hidden Markov Model (HMM) analysis to spike trains recorded from neuronal ensembles in the gustatory cortex (GC) revealed that taste-evoked activity progresses through a sequence of metastable states \cite{Jones2007}. Each metastable state can be described as a pattern of ensemble activity (a collection of firing rates) lasting tens to hundreds of milliseconds. Transitions between different states are characterized by coordinated changes in firing rates across multiple neurons in an ensemble. This description of taste-evoked activity in GC has been successful in capturing essential features of gustatory codes, such as trial-to-trial variability \cite{Jones2007} and learning-dependent changes \cite{MoranKatz2014}. In addition, describing neural responses as sequences of metastable states has been proposed as an explanation for the role of GC in ingestive decisions \cite{MillerKatz2010}. The observation that similar dynamics exist also in frontal \cite{Abeles1995a,Seidemann1996,Durstewitz2010}, motor \cite{Kemere2008}, premotor, and somatosensory \cite{PonceAlvarez2012} cortices further emphasizes the generality of this framework. 
While the functional significance of these transitions among metastable states is being actively investigated \cite{MoranKatz2014}, very little is known about their genesis. A recent attractor-based model has successfully demonstrated that external inputs can give rise to transitions of metastable states by perturbing the network away from a stable spontaneous state \cite{MillerKatz2010}. However, sequences of metastable states could also occur spontaneously, hence reflecting intrinsic network dynamics \cite{DecoHugues2012,DecoJirsa2012,LitwinKumarDoiron2012,MattiaSanchezVives2012,LitwinKumarDoiron2014}. Intrinsic activity patterns have indeed been observed in vivo during periods of ongoing activity, i.e., neural activity in the absence of overt sensory stimulation. Such ongoing patterns have been reported in visual \cite{Arieli1996,Tsodyks1999,Kenet2003,Fiser2004}, somatosensory \cite{Petersen2003,Luczak2007} and auditory \cite{Luczak2009} cortices, and the hippocampus \cite{FosterWilson2006,DibaBuzsaki2007,Luczak2009}. The functional significance of ongoing activity, its origin and its relationship to evoked activity remain, however, elusive. 
Here, we analyzed ongoing activity to investigate whether transitions among metastable states can occur in the absence of sensory stimulation in the GC of alert rats. We found that ongoing activity, much like evoked activity, is characterized by sequences of metastable states. Transitions among consecutive states are triggered by the co-activation of several neurons in the ensemble, with some neurons capable of producing multiple Ð i.e., 3 or more Ð firing rates across the different states (a feature hereby referred to as multi-stability). To elucidate how a network could intrinsically generate transitions among multi-stable states, we introduce a spiking neuron model of GC capable of multi-stable states and exhibiting the same ongoing transition dynamics observed in the data. We prove the existence of multi-stable states analytically via mean field theory, and the existence of transient dynamics via computer simulations of the same model. The model reproduced many properties of the data, including the pattern of correlations between single neurons and ensemble activity, the multi-stability of single neurons, its reduction upon stimulus presentation, and the stimulus-induced reduction of trial-by-trial variability. To our knowledge, this is the first unified and mechanistic model of ongoing and evoked cortical activity.

\section{Materials and Methods}
\label{methods}

\subsection{Experimental subjects and surgical procedures} 

Adult female Long Evans rats (weight: $275-350$ gr) were used for this study \cite{samuelsen2012effects}. Briefly, animals were kept in a 12h:12h light/dark cycle and received ad lib. access to food and water, unless otherwise mentioned. Movable bundles of sixteen microwires (25 $\mu$m nichrome wires coated with fromvar) attached to a Ômini-microdriveÕ \cite{FontaniniKatz2006,samuelsen2012effects} were implanted in GC (AP $1.4$, ML $\pm 5$ from bregma, DV $Ð4.5$ from dura) and secured in place with dental acrylic. After electrode implantation, intra-oral cannulae (IOC) were inserted bilaterally and cemented \cite{PhillipsNorgren1970,FontaniniKatz2005}. At the end of the surgery a positioning bolt for restraint was cemented in the acrylic cap. Rats were given at least 7 days for recovery before starting the behavioral procedures outlined below. All experimental procedures were approved by the Institutional Animal Care and Use Committee of Stony Brook University and complied with University, state, and federal regulations on the care and use of laboratory animals. More details can be found in \cite{samuelsen2012effects}.

\subsection{Behavioral training}
Upon completion of post-surgical recovery, rats began a water restriction regime in which water was made available for $45$ minutes a day. Body-weight was maintained at $>85\%$ of pre-surgical weight. Following few days of water restriction, rats began habituation to head-restraint and training to self-administer fluids through IOCs by pressing a lever. Upon learning to press a lever for water, rats were trained that water-reward was delivered exclusively by pressing the lever following an auditory cue (i.e., a $75$ dB pure tone at a frequency of $5$ kHz). The interval at which lever-pressing delivered water was progressively increased to $40 \pm 3$ s (ITI). In order to receive fluids rats had to press the lever within a $3$ s long window following the cue. Lever-pressing triggered the delivery of fluids and stopped the auditory cue. During training and experimental sessions additional tastants were automatically delivered at random times near the middle of the ITI, at random trials and in the absence of the anticipatory cue. Upon termination of each recording session the electrodes were lowered by at least $150 \mu$m so that a new ensemble could be recorded. Stimuli were delivered through a manifold of 4 fine polyimide tubes slid into the IOC. A computer-controlled, pressurized, solenoid-based system delivered $\sim40\,\mu$l of fluids (opening time $\sim40$ ms) directly into the mouth. The following tastants were delivered: $100$ mM NaCl, $100$ mM sucrose, 100 mM citric acid, and 1 mM quinine HCl. Water ($\sim50\,\mu$l) was delivered to rinse the mouth clean through a second IOC five seconds after the delivery of each tastant. Each tastant was delivered for at least 6 trials in each condition. 

\subsection{Electrophysiological recordings} 
Single neuron action potentials were amplified, bandpass filtered (at $300-8000$ Hz), digitized and recorded to a computer (Plexon, Dallas, TX). Single units of at least 3:1 signal-to-noise ratio were isolated using a template algorithm, cluster cutting techniques and examination of inter-spike interval plots (Offline Sorter, Plexon, Dallas, TX). 

\subsection{Data analysis} 
All data analysis and model simulations were performed using custom software written in Matlab (Mathworks, Natick, MA, USA), Mathematica (Wolfram Research, Champaign, IL), and C. Starting from a pool of $299$ single neurons in $37$ sessions, neurons with peak firing rate lower than 1 Hz (defined as silent) were excluded from further analysis, as well as neurons with a large peak around the $6-10$ Hz in the spike power spectrum, which were considered somatosensory \cite{Katz2001,samuelsen2012effects,HorstLaubach2013}. Only ensembles with more than two simultaneously recorded neurons were included in the rest of the analyses ($27$ ensembles with $167$ neurons). We analyzed ongoing activity in the 5 seconds interval preceding either the auditory cue or taste delivery, and evoked activity in the 5 seconds interval following taste delivery exclusively in trials without anticipatory cue.

\subsubsection{Hidden Markov Model (HMM) analysis}
The details of the HMM have been reported in e.g. \cite{Abeles1995a,Jones2007,Escola2011,PonceAlvarez2012}; here we briefly outline the methods of this procedure, especially where our methods differ from previous accounts. Under the HMM, a system of N recorded neurons is assumed to be in one of a predetermined number of hidden (or latent) states. Each state is defined as a vector of N firing rates, one for each simultaneously recorded neuron. In each state, the neurons were assumed to discharge as stationary Poisson processes (``Poisson-HMM"). 

We denote by $[y_i (1),\ldots,y_i (T)]$ the spike train of the $i$-th neuron in the current trial, where $y_i (t)=k$ if $k$ spikes occur in the interval $[t,t+dt]$ and zero otherwise; $dt=1$ ms and $T$ is the total duration of the trial in ms (or, alternatively, the number of bins since $dt=1$ ms).  Denoting with $S_t$ the hidden state of the ensemble at time $t$, the probability of having $k$ spikes/s from neuron $i$ in a given state $m$ in the interval $dt$ is given by the Poisson distribution:
\be
p\left(y_i (t)=k | S_t=m\right)={(\nu_i (m)dt)^k e^{-\nu_i (m)dt}\over k!} \, ,
\ee
where $\nu_i (m)$ is the firing rate of neuron $i$ in state $m$. The firing rates $\nu_i (m)$ completely define the states and are also called Ôemission probabilitiesÕ in HMM parlance. We matched the HMM model to the data segmented in $1$ ms bins. Given the short duration of the bins, we assumed that in each bin either zero or one spikes are emitted by each neuron, with probabilities $p\left(y_i (t)=0 | S_t=m\right)=e^{-\nu_i (m)\cdot dt}$ and $p\left(y_i (t)=1|S_t=m\right)=1-e^{-\nu_i (m)\cdot dt}$, respectively. We also neglected the simultaneous firing of two or more neurons (a rare event): if more than one neuron fired in a given bin, a single spike was randomly assigned to one of the firing neurons. 

The HHM model is fully determined by the emission probabilities defining the states and the transition probabilities among those states (assumed to obey the Markov property). The emission and transition probabilities were found by maximization of the log-likelihood of the data given the model via the expectation-minimization (EM), or Baum-Welch, algorithm \cite{Rabiner1989}. This is known as ``training the HMM." For each session and type of activity (ongoing vs. evoked), ensemble spiking activity from all trials was binned at $1$ ms intervals prior to training assuming a fixed number of latent states $M$ \cite{Jones2007,Escola2011}. We used always the same state as the initial state (e.g., state ``$1"$), but ran the algorithm starting $200$ ms prior to the interval of interest to obtain an unbiased estimate of the first state of each sequence. For each given number of states $M$, the Baum-Welch algorithm was run $5$ times, each time with random initial conditions for the transition and emission probabilities. The procedure was repeated for $M$ ranging from $10$ to $20$ for ongoing activity trials, and from $10$ to $40$ for evoked activity trials, based on previous accounts \cite{Jones2007,MillerKatz2010,Escola2011}. For evoked activity, each HMM was trained on all four tastes simultaneously, using overall the same number of trials as for ongoing activity. Of the models thus obtained, the one with largest total likelihood was taken as the best HMM match to the data, and then used to estimate the probability of the states given the model and the observations in each bin of each trial (a procedure known as ``decoding"). During decoding, only those states with probability exceeding $80\%$ in at least $50$ consecutive $1$-ms bins were retained (henceforth denoted simply as ``states"). 

For the firing rate analysis of Fig. 2, the firing rates $\nu_i (m)$ were obtained from the analytical solution of the maximization step of the Baum-Welch algorithm, 
\be
\nu_i (m)=-{1\over dt} \ln\left(1-{\sum_{t=1}^T r_m(t)y_i(t) \over \sum_{t=1}^T r_m(t)}\right)  \, .
\ee
Here, $r_m (t)=p\left(S_t=m | y(1),\ldots,y(T)\right)$ is the probability that the state at time $t$ is $m$, given the observations. Note that this procedure allows for sampling the variability of $\nu_i (m)$ across trials, which can be then used for inferential statistics (see below, ``State specificity and firing rate distributions").

The distribution of state durations across sessions, $t_s$, was fitted with an exponential function $f(t_s;a,b)=a\cdot \exp\,(b\, t_s)$ using non-linear least squares minimization and $95\%$ confidence intervals (CIs) were computed according to a standard procedure (see e.g. \cite{Wolberg2006}, p. 50).  

\subsubsection{Firing rate modulations and change-point analysis}
We detected changes in single neurons' instantaneous firing rate (in single trials) by using the change point procedure based on the cumulative distribution of spike times of \cite{Gallistel2004}. Briefly, the procedure selects a sequence of putative variations of firing rate, called ``change points," as the points where the cumulative record of the spike count produces local maximal changes in slope. A putative change point was accepted as valid if the spike count rates before and after the change point were significantly different (binomial test, $p<0.05$). See \cite{Gallistel2004} for details. This method, which is based on a binomial test for the spike count \cite{Gallistel2004}, was chosen because it is a simple and general tool for analyzing cumulative records, and had been already empirically validated for the analysis of change points in GC \cite{Jezzini2013}.

\subsubsection{Single neuron responsiveness to stimuli}
Neurons with significant post-stimulus modulations of firing rate were defined as stimulus-responsive. The analysis was based on the CP procedure performed in $[-1,5]$ s intervals around stimulus delivery occurring at time zero (see \cite{Jezzini2013} for further details). The CP procedure was performed on the cumulative record of spike counts over all evoked trials for a given taste. If any CP was detected before taste delivery, the CP analysis was repeated with a more stringent criterion, until no pre-stimulus change points were detected. If no CP was found in response to any of the four tastes, the neuron was deemed not responsive. If a CP was detected in the post-delivery interval for at least one taste, a t-test was performed to establish whether the post-CP activity was significantly different from baseline activity during the $1$ s interval before taste delivery ($p < 0.05$), and only then accepted as a significant CP. Neurons with at least one significant CP were considered stimulus-responsive. 

\subsubsection{Change point-triggered average (CPTA)}

The CPTA of ensemble transitions, formally reminiscent of the spike-triggered average \cite{Chichilnisky2001,Dayan2001}, estimates the average HMM transition at a lag $t$ from a CP. Given a set of CPs at times $ [t_1^{(i)},\ldots,t_P^{(i)}]$ for the $i$-th neuron, its CPTA is defined as 
\be
\textrm{CPTA}^{(i)}(t)=\left\langle{1\over P}\sum_{p=1}^P I_{[t+t_p^{(i)},t+t_p^{(i)}+\Delta t]} \right\rangle \ ,
\ee
where $\left\langle\ldots\right\rangle$ denotes average across trials, $\Delta t=50$ ms, and $I_{[a,b]} =1$ if a transition occurs in the interval $[a,b]$, and zero otherwise. Significance of the CPTA was assessed by comparing the simultaneously recorded data with $100$ trial-shuffled ensembles (t-test, $p<0.05$, Bonferroni corrected for multiple bins), in which we randomly permuted each spike train across trials, in order to scramble the simultaneity of the ensemble recordings. A CPTA peak at zero indicates that, on average, CPs co-occur with ensemble transitions, while e.g. a significant CPTA for positive lag indicates that CPs tend to precede transitions (see Fig. 2B).

\subsubsection{State specificity and firing rate distributions}

We defined a neuron as ``state-specific" if its firing rate varied across different HMM states. In order to assess state specificity for the $i$-th unit, we collected the set of firing rates across all trials and states (see earlier Section ``Hidden Markov Model (HMM) analysis"). To determine whether the distribution of firing rates varied across states, we performed a non-parametric one-way ANOVA (unbalanced Kruskal-Wallis, $p<0.05$). The smallest number of significantly different firing rates for a given neuron in a given state was found with a post-hoc multiple comparison rank analysis (with Bonferroni correction). Given a p-value $p_{lk}$ for the pairwise post-hoc comparison between states $l$ and $k$, we considered the symmetric $M\times M$ matrix $A$ with elements $A_{lk}=1$ if the rates were different ($p_{lk}<0.05$) and $A_{lk}=0$ otherwise. For example, consider the case of $4$ states and the following two outcomes of multiple comparisons: 
\be
A^{(1)}=
\left[
\begin{array}{cccc}
\cdot &0 & 1 & 1 \\
\cdot &\cdot & 1 & 1 \\
\cdot &\cdot & \cdot & 1 \\
\cdot &\cdot & \cdot & \cdot 
\end{array}
\right] \ ,\qquad
A^{(2)}=
\left[
\begin{array}{cccc}
\cdot &0 & 0 & 1 \\
\cdot &\cdot & 0 & 1 \\
\cdot &\cdot & \cdot & 0 \\
\cdot &\cdot & \cdot & \cdot 
\end{array}
\right] \ .
\ee
In the first case, the firing rate in state 1 is different than in states $3$ and $4$ (first row); it is different in state $2$ vs. $3$ and $4$ (second row), and in state $3$ vs. $4$ (third row). The conclusion is that $f_1\neq f_3, \, f_1\neq f_4$ (from the first row), and $ f_3\neq f_4$ (from the 3rd row), while $f_2$, although different from $f_3$ and $f_4$, may not be different from  $f_1$. It follows that at least $3$ different firing rates are found across states. In the example of matrix $A^{(2)}$, the smallest number of different rates is $2$ (e.g., $f_1\neq f_4$).

\subsubsection{Single unit-ensemble correlations and $\phi$ statistics}

We estimated the nonparametric correlation between ensemble and single units firing rate modulations in single trials, by means of their $\phi$ statistics. For each ensemble transition between consecutive states $m\to n$, we looked at whether the $i$-th neuron increased $(s^i (m\to n)=+1)$ or decreased $(s^i (m\to n)=-1)$ its firing rate during the transition, and analogously for the ensemble mean firing rate $s^{\textrm{ens}} (m\to n)=\pm 1$. For each transitions $m\to n$, we counted the number of ensemble neurons that increased their firing rate when the whole ensemble increased its firing rate
\be
n_{++} (m\to n)=\sum_{i=1}^N\left(s^i (m\to n)=+1  \,\,\&\,\,  s^{\textrm{ens}} (m\to n)=+1\right) \ , 
\ee
and analogously for the other three possibilities $n_{+-},\,n_{-+},\,n_{--}$, thus obtaining a $2\times 2$ matrix $n_{ab} (m\to n)$, for $a,b=+,-$. We then compiled the contingency matrix with elements $N_{ab}=\sum n_{ab} (m\to n)$, by summing each $n_{ab} (m\to n)$ over all transitions and sessions. Based on these quantities, the $\phi$-statistics \cite{Pearson1904,Cramer1946} was computed as 
\be
\phi={\det(N_{ab})\over  \sqrt{(\sum_bN_{+b})(\sum_bN_{-b})(\sum_aN_{a+})(\sum_aN_{a-})}} \ .
\ee
If changes in single neurons perfectly matched ensemble changes (as in the case of UP and DOWN states), then $\phi=1$; while $\phi=0$ in case of no correlation and $0<\phi<1$ in case of partial co-activation of neurons and ensembles.

\subsubsection{Fano factor}
The raw Fano factor (FF) for the $i$-th unit is defined as
\be
F={\textrm{var}[N_i (t)]\over\langle N_i (t)\rangle} \  ,
\ee
where $N_i (t)$ is the spike count for the  $i$-th neuron in the $[t,t+\Delta t]$ bin, and $\langle\ldots,\rangle$ and var are the mean and variance across trials. In order to control for the difference in firing rates at different times, the raw FF was mean-matched across all bins and conditions, and its mean and $95\%$ CI were estimated via linear regression (see \cite{Churchland2010} for details). We used a window of $200$ ms for the comparison between model and data. Results were qualitatively similar for a wide range of windows sizes ($\Delta t =50$ to $500$ ms), sliding along at $50$ ms steps. 

\subsection{Spiking neuron network model}
Our recurrent network model comprised $N=5000$ randomly connected leaky integrate-and-fire (LIF) neurons, with a fraction $n_E=0.8$ of excitatory (E) and $n_I=0.2$ inhibitory (I) neurons. Connection probability $p_{\beta\alpha}$ from neurons in population $\alpha\in E,I$ to neurons in population $\beta\in E,I$ were $p_{EE}=0.2$ and $p_{EI}=p_{IE}=p_{II}=0.5$. A fraction $f=0.9$ of excitatory neurons were arranged into $Q$ different clusters, with the remaining fraction belonging to an unstructured (``background") population \cite{Amit1997b}. Synaptic weights $J_{\beta\alpha}$ from neurons in population  $\alpha\in E,I$ to neurons in population $\beta\in E,I$  scaled with $N$ as $J_{\beta\alpha}=j_{\beta\alpha}/\sqrt{N}$, with $j_{\beta\alpha}$ constants drawn from a normal distribution with the following mean values (units of mV): $j_{EI}=3.18,\,j_{IE}=1.06,\,j_{II}=4.24,\,j_{EE}=1.77$; and variance $\textrm{Var}[J]=\delta^2J^2$, with $\delta^2=0.01$. Within an excitatory cluster synaptic weights were potentiated, i.e. they took values $J_+ j_{EE}$ with $J_+>1$, while synaptic weights between neurons belonging to different clusters were depressed to values $J_- j_{EE}$, with $J_-=1-\gamma (J_+-1)f/Q<1$, with $\gamma=0.5$. The latter relationship between $J_+$ and $J_-$ insures balance between overall potentiation and depression in the network \cite{Amit1997b}. 
Below spike threshold, the membrane potential $V$ of each LIF neuron evolved according to 
\be
\tau_m  {dV\over dt}=-V+\tau_m (I_{\textrm{rec}}+I_\textrm{ext}+I_\textrm{stim} ) \  ,
\ee
with a membrane time constant $\tau_m=20$ ms for excitatory and $10$ ms for inhibitory neurons. The input current was the sum of a recurrent input $I_\textrm{rec}$, an external current $I_\textrm{ext}$ representing an ongoing afferent input from other areas, and an input stimulus $I_\textrm{stim}$ representing a delivered taste during evoked activity only. In our units, a membrane capacitance of $1$ nF is set to $1$. A spike was said to be emitted when $V$ crossed a threshold $V_\textrm{thr}$, after which $V$ was reset to a potential $V_\textrm{reset}=0$ for a refractory period of $\tau_\textrm{ref}=5$ ms. Spike thresholds were chosen so that, in the unstructured network (i.e., with $J_+=J_-=1$), the E and I populations had average firing rates of $3$ and $5$ spikes/s, respectively \cite{Amit1997b}. The recurrent synaptic input $I_\textrm{rec}^i$ to neuron i evolved according to the dynamical equation
\be
\tau_s  {dI_\textrm{rec}^i\over dt}=-I_\textrm{rec}^i+\sum_{j=1}^NJ_{ij}  \sum_k\delta(t-t_k^j )  \  ,
\ee
where $t_k^j$ was the arrival time of $k$-th spike from the $j$-th pre-synaptic neuron, and $\tau_s$ was the synaptic time constant ($3$ and $2$ ms for E and I neurons, respectively). The PSC elicited by a single incoming spike was ${J_{ij}\over \tau_s}  e^{-t/\tau_s}\Theta(t)$, where $\Theta(t)=1$ for $t\geq0$, and $\Theta(t)=0$ otherwise. The ongoing external current to a neuron in population $\alpha$ was constant and given by
\be
I_\textrm{ext}=N_\textrm{ext} p_{\alpha0} J_{\alpha0} \nu_\textrm{ext} \ ,
\ee
where $N_\textrm{ext}=n_E N,\, p_{\alpha0}=p_{EE}, \, J_{\alpha0}=j_{\alpha0}/\sqrt{N}$ with $j_{E0}=0.3, \, j_{I0}=0.1$, and $\nu_\textrm{ext}=7$ spikes/s. During evoked activity, stimulus-selective neurons received an additional transient input representing one of the four incoming stimuli. The percentage of neurons responsive to one, two, three or four stimuli was modeled after the estimates obtained from the data, which implied that stimuli targeted overlapping neurons (see Results). We tested two alternative model stimuli: a biologically realistic stimulus $\nu_\textrm{stim}^\textrm{th} (t)$ resembling thalamic stimulation \cite{Liu2014}, modeled as a double exponential with peak amplitude of $0.3\nu_\textrm{ext}$ and rise times of $50$ ms and decay times of $500$ ms, or a stimulus of constant amplitude $\nu_\textrm{stim}^\textrm{box}$ ranging from $0$ to $0.5\nu_\textrm{ext}$ (``box" stimulus). In the following we measure the stimulus amplitude as percentage of $\nu_\textrm{ext}$ (e.g., ``$30\%$" corresponds to $\nu_\textrm{stim}=0.3 \nu_\textrm{ext}$. The onset of each stimulus was always $t=0$, the time of taste delivery. The stimulus current to a neuron in population $\alpha$ was constant and given by
$I_\textrm{stim}=N_\textrm{ext} p_{\alpha0} J_{\alpha0} \nu_\textrm{stim}$.

\subsubsection{Mean field analysis of the model}
The spiking network model described in the previous subsection is a complex system capable of many behaviors depending on the parameter values. One main aim of the model is finding under what conditions it can sustain multiple configurations of activity that can be later interpreted as HMM states. Parameter search was used relying on an analytical procedure for networks of LIF neurons known as ``mean field theory" or ``population density approach" (see e.g. \cite{Amit1997b,Brunel1999,Fusi1999}). Under the conditions stated below, this theory provides a global picture of network behavior together with the associated parameter values, which can then be tested in model simulations. 
Under typical conditions, each neuron of the network receives a large number of small post-synaptic currents (PSCs) per integration time constant. In such a case, the dynamics of the network can be analyzed under the diffusion approximation, which is amenable to the population density approach. The network has $\alpha=1,É,Q+2$ sub-populations, where the first $Q$ indices label the $Q$ excitatory clusters, $\alpha=Q+1$ labels the ``background" excitatory population, and $\alpha=Q+2$ labels the homogeneous inhibitory population. In the diffusion approximation \cite{Tuckwell1988,Lansky1999,Richardson2004}, the input to each neuron is completely characterized by the infinitesimal mean $\mu_\alpha$   and variance $\sigma_\alpha^2$ of the post-synaptic potential. Adding up the contributions from all afferent inputs, $\mu_\alpha$ and $\sigma_\alpha^2$ for an excitatory neuron belonging to cluster $\alpha$ are given by \cite{Amit1997b}
\bea
\mu_\alpha&=&\tau_{m,E} \sqrt{N} \Bigl({n_E f\over Q} \left(p_{EE} J_+ j_{EE} \nu_\alpha+\sum_{\beta=1}^{Q-1}p_{EE}  J_- j_{EE} \nu_\beta \right)+n_E (1-f) p_{EE} J_- j_{EE} \nu_E^{(bg)}\nn\\
&&-n_I p_{EI} j_{EI} \nu_I+n_E p_{E0} j_{E0} \nu_\textrm{ext} \Bigr) \  ,  \nn \\
\sigma_\alpha^2&=&\tau_{m,E}  \Bigl({n_E f\over Q} \left(p_{EE} (J_+ j_{EE} )^2(1+\delta^2) \nu_\alpha+\sum_{\beta=1}^{Q-1}p_{EE}  (J_- j_{EE} )^2(1+\delta^2) \nu_\beta \right)\nn\\
&&+n_E (1-f) p_{EE} (J_- j_{EE} )^2(1+\delta^2) \nu_E^{(bg)}+n_I p_{EI} j_{EI}^2(1+\delta^2) \nu_I \Bigr)  \ , \nn
\eea
where $\nu_E^{(bg)}$ is the firing rate of the unstructured (``background") E population. Afferent current and variance to a neuron belonging to the background E population and to the homogeneous inhibitory population are
\bea
\mu_E^{(bg) }&=&\tau_{m,E} \sqrt{N} \Bigl({n_E f \over Q} \sum_{\beta=1}^Qp_{EE}  J_- j_{EE} \nu_\beta+n_E (1-f) p_{EE} j_{EE} \nu_E^{(bg)}\nn\\
&&-n_I p_{EI} j_{EI} \nu_I+n_E p_{E0} j_{E0} \nu_\textrm{ext} \Bigr)  \ , \nn \\
(\sigma_E^{(bg) } )^2&=&\tau_{m,E} \Bigl({n_E f\over Q} \sum_{\beta=1}^Q p_{EE}  (J_- j_{EE} )^2(1+\delta^2) \nu_\beta+n_E (1-f) p_{EE} j_{EE}^2(1+\delta^2) \nu_E^{(bg)}\nn\\
&&+n_I p_{EI} j_{EI}^2(1+\delta^2) \nu_I \Bigr) \ ,\nn\\
\mu_I&=&\tau_{m,I} \sqrt{N} \Bigl({n_E f \over Q} \sum_{\beta=1}^Qp_{IE}  j_{IE} \nu_\beta+n_E (1-f) p_{IE} j_{IE} \nu_E^{(bg)}\nn\\
&&-n_I p_{II} j_{II} \nu_I+n_E p_{I0} j_{I0} \nu_\textrm{ext} \Bigr) \  ,\nn\\ 
\sigma_I^2&=&\tau_{m,I} \Bigl({n_E f\over Q} \sum_{\beta=1}^Qp_{IE}  j_{IE}^2(1+\delta^2) \nu_\beta+n_E (1-f) p_{IE} j_{IE}^2(1+\delta^2) \nu_E^{(bg)}\nn\\
&&+n_I p_{II} j_{II}^2(1+\delta^2) \nu_I \Bigr).
\eea
Parameters were chosen so as to have a balanced unstructured network. In other words, our network with $J_+=J_-=1$ (where all E$\to$E synaptic weights are equal) would operate in the balanced asynchronous regime where incoming contributions from excitatory and inhibitory inputs balance out and the peri-stimulus time interval over time is approximately flat. In such a regime, one can solve for the excitatory and inhibitory firing rates as linear functions of the external firing rate $\nu_\textrm{ext}$, up to terms of ${\cal O}(1/\sqrt{N})$, provided the connection probabilities and synaptic weights satisfy one of the following conditions \cite{van1996chaos,vreeswijk1998chaotic,Renart2010}: either
\be
j_{II}>{p_{EI}  p_{IE}  j_{EI}  j_{IE}\over p_{EE}  p_{II}  j_{EE} }  \ ,\qquad  j_{I0}>{p_{II}  j_{E0}  j_{II}\over p_{EI} j_{EI} } \ ,
\ee
or
\be
j_{II}<{p_{EI}  p_{IE}  j_{EI}  j_{IE}\over p_{EE}  p_{II}  j_{EE} }  \ , \qquad j_{I0}<{p_{II}  j_{E0}  j_{II}\over p_{EI} j_{EI} } \  .
\ee
Asynchronous activity was confirmed by computing the network synchrony, defined as \cite{Golomb2000}
\be
\Sigma=\sqrt{\sigma_\textrm{pop}^2\over [\sigma_i^2] } \ ,
\ee
where $\sigma_i^2$ is the time-averaged variance in the instantaneous firing rate $\nu_i (t)$ of the $i$-th neuron, $[\ldots]$ denotes the average over all neurons in the network, and $\sigma_\textrm{pop}^2$ is the time-averaged variance of the instantaneous population firing rate $[\nu_i (t)]$. The network synchrony $\Sigma$ is expected to be of ${\cal O}(1)$ in a fully synchronized network, but of order ${\cal O}(1/\sqrt{N})$ in networks of size $N$ that are in the asynchronous state. We used bins of variable size (from $10$ ms to $50$ ms) to estimate $\Sigma$, obtaining similar results in all cases.
The unstructured network has only one dynamical state, i.e., a stationary point of activity where all E and I neurons have constant firing rate $\nu_E$ and $\nu_I$, respectively. In the structured network (where $J_+>1$), the network undergoes continuous transitions among a repertoire of states, as shown in the main text. Not to confuse the networkÕs states of activity with the HMM states, we shall from now on use the term ``network configurations" instead. Admissible network configurations must satisfy the $Q+2$ self-consistent mean field equations \cite{Amit1997b}
\be 
\nu_\alpha=F_\alpha \left(\mu_\alpha (\overrightarrow{\nu}),\sigma_\alpha^2 (\overrightarrow{\nu})\right)  \ ,
\ee
where $\overrightarrow{\nu}=[\nu_1,\ldots,\nu_Q,\nu_E^{(bg)},\nu_I ]$ is the firing rate vector and $F_\alpha (\mu_\alpha,\sigma_\alpha^2 )$ is the current-to-rate response function of the LIF neurons. For fast synaptic times, i.e. $\tau_s/\tau_m <<1$, $F_\alpha (\mu_\alpha,\sigma_\alpha^2 )$ is well approximated by \cite{Brunel1998,Fourcaud2002}
\be
F_\alpha (\mu_\alpha,\sigma_\alpha )=\left(\tau_\textrm{ref}+\tau_{m,\alpha}\sqrt{\pi}\int_{H_{\textrm{eff},\alpha}}^{\Theta_{\textrm{eff},\alpha}}e^{u^2}  [1+\textrm{erf}(u)]\right)^{-1},
\ee
where 
\bea
\Theta_{\textrm{eff},\alpha}&=&{V_{\textrm{thr},\alpha}-\mu_\alpha\over \sigma_\alpha} +ak_\alpha \ ,\nn\\
H_{\textrm{eff},\alpha}&=&{V_{\textrm{reset},\alpha}-\mu_\alpha\over \sigma_\alpha} +ak_\alpha,\nn
\eea
where $k_\alpha=\sqrt{\tau_{s,\alpha}/\tau_{m,\alpha} }$ is the square root of the ratio of synaptic time constant to membrane time constant, and $a=|\zeta(1/2)|/\sqrt{2}\sim1.03$. This theoretical response function has been fitted successfully to the firing rate of neocortical neurons in the presence of in vivo-like fluctuations \cite{Rauch2003,LaCamera2006,LaCamera2008}. 
The fixed points $\overrightarrow{\nu}^*$ of the mean field equations were found with NewtonÕs method \cite{Press2007}. The fixed points can be either stable (attractors) or unstable depending on the eigenvalues $\lambda_\alpha$ of the stability matrix
\be
S_{\alpha\beta}={1\over \tau_{s,\alpha}}\left({\partial F_\alpha \left(\mu_\alpha (\overrightarrow{\nu}),\sigma_\alpha^2 (\overrightarrow{\nu})\right)\over \partial\nu_\beta }-{\partial  F_\alpha \left(\mu_\alpha (\overrightarrow{\nu}),\sigma_\alpha^2 (\overrightarrow{\nu})\right)\over \partial\sigma_\alpha^2 }  {\partial\sigma_\alpha^2\over \partial\nu_\beta }-\delta_{\alpha\beta} \right) \  ,
\label{stabilitymatrix}
\ee
evaluated at the fixed point $\overrightarrow{\nu}^*$ \cite{Mascaro1999}.\footnote{Equal indices are not summed over. The published version of the current article in \cite{mazzucato2015dynamics} has a typo that was corrected in \ref{stabilitymatrix}.} If all eigenvalues have negative real part, the fixed point is stable (attractor). If at least one eigenvalue has positive real part, the fixed point is unstable. Stability is meant with respect to an approximate linearized dynamics of the mean and variance of the input current:
\bea
\tau_{s,\alpha}  {dm_\alpha\over dt}&=&- m_\alpha+\mu_\alpha (\overrightarrow{\nu}) \ ,\nn\\
{\tau_{s,\alpha}\over 2}  {ds_\alpha^2\over dt}&=&- s_\alpha^2+\sigma_\alpha^2 (\overrightarrow{\nu}) \ , \nn\\
\nu_\alpha (t)&=&F_\alpha \left(m_\alpha (\overrightarrow{\nu}),s_\alpha^2 (\overrightarrow{\nu})\right) \ , \nn
\eea
where $\mu_\alpha$ and $\sigma_\alpha^2$ are the stationary values for fixed $\overrightarrow{nu}$ given earlier. For fast synaptic dynamics in the asynchronous balanced regime, these rate dynamics are in very good agreement with simulations (\cite{LaCamera2004} Ð see \cite{Renart2004,Giugliano2008} for more detailed discussions). 

\subsubsection{Model simulations}

The dynamical equations of the LIF neurons were integrated with the Euler algorithm with a time step of $dt=0.1$ ms. We simulated thirty ensembles during ongoing activity and twenty ensembles during evoked activity. We chose four different stimuli per session during evoked activity (as in the data). Trials were $5$ seconds long. In all cases, we pre-processed each ensemble by picking at random a distribution of neurons per ensemble that matched the distribution of ensemble sizes found in the data, and performed the same analyses performed on the data (i.e., HMM, CPTA, FF, and so on). The range of hidden states $M$ for the HMM in the model was the same as the one used for the data (see above), except in the case of analyses involving simulated ensembles of $30$ neurons (Fig. 5A), where $M$ varied from $20$ to $60$ states. To determine the number of active clusters at any given time (Fig. 4), a cluster was considered active if its population firing rate was higher than $20$ spikes/s in a $50$ ms bin. This criterion gave a good separation of the firing rates of active and inactive clusters in the multi-stable region (Fig. 3B, region with $J_+>J'$). A cluster was active in a particular bin if its population firing rate crossed the $20$ spks/s threshold in that bin.

\section{Results}
\label{results}

\begin{figure}[ht]
\begin{center}
\hspace*{-0.5cm}                                                           
\includegraphics[width=1.04\textwidth]{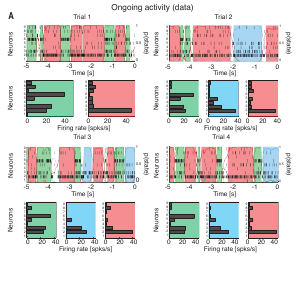}
\end{center}
\vspace*{-2.2cm}
\caption{{Ongoing activity is characterized by sequences of states}. Four representative trials from one ensemble of nine simultaneously recorded neurons, segmented according to their ensemble states (HMM analysis). Thin black vertical lines are action potentials. States are color-coded; smooth colored lines represent the probability for each state; shaded colored areas indicate intervals where the probability of a state exceeds $80\%$. Below each single-trial population raster, average firing rates across simultaneously recorded neurons are plotted (states are color-coded). $X$-axis for population rasters: time preceding the next event at ($0 =$ stimulus delivery); $X$-axis for average firing rates panels: firing rates (spks/s); $Y$-axis for population rasters: left, ensemble neuron index, right, probability of HMM states; Y-axis for firing rate panels: ensemble neuron index.}
\label{Fig1}
\end{figure}

\begin{figure}[ht]
\vspace{-1cm}                                                        
\begin{center}
\hspace*{-0.4cm}                                                           
\includegraphics[width=1.03\textwidth]{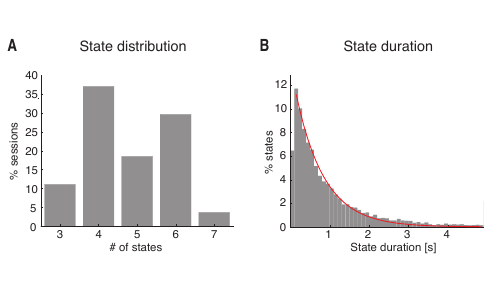}
\end{center}
\vspace*{-1.5cm}
\caption{{\it Features of ongoing activity states}. {\bf A}: distribution of number of states per ensemble across all sessions (only states lasting longer than $50$ ms are included). $X$-axis: number of states per ensemble; $Y$-axis: frequency of occurrence. {\bf B}: distribution of state durations across all sessions together with the least squares exponential fit (red line; see Appendix~\ref{methods}). $X$-axis: state duration; $Y$-axis: frequency of states occurrence.}
\label{Fig2}
\end{figure}

\subsection{Characterization of ongoing cortical activity as a sequence of ensemble states}

We analyzed ensemble activity from extracellular recordings in the gustatory cortex (GC) of behaving rats. In our experimental paradigm, rats were delivered one out of four tastants (sucrose, sodium chloride, citric acid and quinine, denoted respectively as S, N, C, and Q) through an intra-oral cannula, followed by a water rinse to clean the mouth from taste residues \cite{samuelsen2012effects}. In half of the trials, taste delivery was preceded by an auditory cue ($75$ db, $5$ kHz), signaling taste availability upon lever press; in the other half of the trials, taste was delivered randomly in the absence of an anticipatory cue. Expected and unexpected trials were randomly interleaved. Rats were engaged in a task to prevent periods of rest. We analyzed ongoing activity in the $5$ seconds interval preceding either the auditory cue or taste delivery, and evoked activity in the $5$ seconds interval following taste delivery in trials without anticipatory cue. The interval length was chosen based on the fact that population activity in GC encodes taste-related information for at least five seconds after taste delivery \cite{Jezzini2013} and the ongoing interval was designed to match the length of the evoked one.
Visual inspection of ensemble spiking activity during ongoing inter-trial periods (Fig. \ref{Fig1}) reveals that several neurons simultaneously change their firing rates. This suggests that, similarly to taste-evoked activity \cite{Jones2007}, ongoing activity in GC could be characterized in terms of sequences of states, where each state is defined as a collection of constant firing rates across simultaneously recorded neurons (Fig. \ref{Fig1}; see Appendix~\ref{methods} for details). We assumed that the dynamics of state transitions occurred according to a Markov stochastic process and performed a hidden Markov model (HMM) analysis in single trials \cite{Seidemann1996,Jones2007,Escola2011,PonceAlvarez2012}. For each neural ensemble and type of activity (ongoing vs. evoked), spike trains were fitted to several Poisson-HMMs that differed for number of latent states and the initial conditions. The model providing the largest likelihood was selected as the best model, but only states with probability larger than $80\%$ in at least $50$ consecutive $1$ ms bins were retained as valid states (called simply ``states" in the following; see Appendix~\ref{methods} for details). The number of states across sessions ranged from $3$ to $7$ (mean$\pm$SD: $4.8\pm1.1$, Fig. \ref{Fig2}A) with approximately exponentially distributed state durations with mean of $717$ ms ($95\%$ CI: $[687, 747]$, Fig. \ref{Fig2}B). The number of states was correlated with the number of neurons in each state ($R^2=0.3,\,p<0.01$). However, such correlation disappeared when rarely occurring states (occurring for less than $3\%$ of the total session duration) were excluded ($R^2=0.1,\,p>0.1$), suggesting that the most frequent states are present even in small ensembles.

\begin{figure}[ht]
\vspace{-1cm}                                                        
\begin{center}
\hspace*{-0.55cm}                                                           
\includegraphics[width=1.05\textwidth]{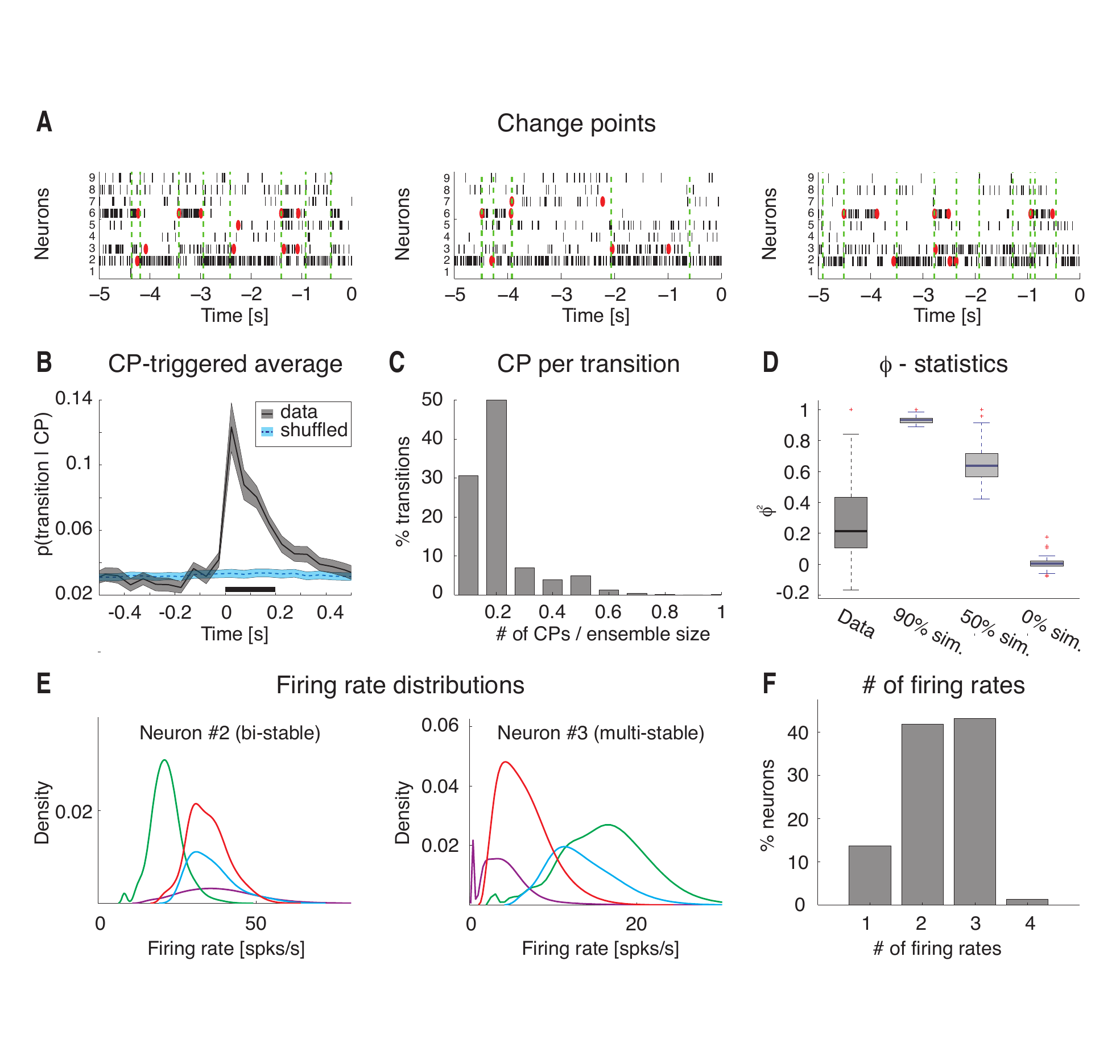}
\end{center}
\vspace*{-1.3cm}
\caption{{\it Single neurons co-activation drives ensemble transitions}. {\bf A}: change points (CPs, red dots) and state transitions (vertical dashed green lines) are superimposed to a raster plot of simultaneously recorded neurons (same trials as the first three in Fig. \ref{Fig1}). $X$-axis for upper panel: time preceding taste delivery ($0 =$ stimulus delivery); $Y$-axis: ensemble neuron index. {\bf B}: CP triggered average (CPTA) of ensemble transitions for the empirical dataset (black) and for the trial-shuffled dataset (blue). Shaded bounds: standard errors. Thick black line below traces: $p<0.05$ (bin-wise t-test with multiple-bin Bonferroni correction). $X$-axis: time lag from CP (s); $Y$-axis: probability of a transition given a CP at $t=0$. {\bf C}: fraction of transitions co-occuring with CPs within a $200$ ms window prior to the state transition (significant window from panel {\bf B}). $X$-axis: number of single neuronÕs CPs divided by the ensemble size; $Y$-axis: fraction of transitions co-occurring with CPs (\%). {\bf D}: correlation ($\phi^2$-statistics) between the signs of single neuronsÕ and ensemble firing rate changes for the empirical (``Data") and simulated datasets. In the latter, $90\%$, $50\%$ and $0\%$ of simulated neurons had firing rate changes matched to those of the whole ensemble (boxes in  box-plots represent $95\%$ CIs). $Y$-axis: $\phi^2$-statistics. {\bf E}: firing rate distributions across states for representative neurons $2$ and $3$ from the ensemble in Fig. \ref{Fig1} and panel {\bf A} above (color-coded as in Fig. \ref{Fig1}, each curve represents the empirical distribution of firing rates in each state). $X$-axis: firing rate (spks/s); $Y$-axis: probability density of states. {\bf F}: number of different firing rates per neuron across all sessions and neurons: $42\%$ of all neurons had $3$ or more different firing rates across states. $X$-axis: minimal number of different firing rates across states; $Y$-axis: fraction of neurons (\%).}
\label{Fig3}
\end{figure}

To gain further insight into the structure revealed by HMM analysis, we pitted the ensemble state transitions against modulations of firing rate in single neurons. The latter were found using a change point (CP) procedure applied to the cumulative distribution of spike counts \cite{Gallistel2004,Jezzini2013}. Examples from three representative trials are shown in Fig. \ref{Fig3}A, where it can be seen that CPs from multiple neurons (red dots) tend to align with ensemble state transitions (dashed vertical green lines). The relationship between CPs and state transitions was characterized with a CP-triggered average (CPTA) of state transitions. The CPTA estimates the likelihood of observing a state transition at time $t$, given a CP in any neuron of the ensemble at time $t=0$. If detection of CPs and state transitions were instantaneous, a significant peak of the CPTA at positive times would indicate that state transitions tend to follow a CP, whereas a peak at negative times would indicate that state transitions tend to precede CPs. We found a significant positive CPTA in the interval between $t=0$ and $t=200$ ms (Fig. \ref{Fig3}B, black trace), with a peak at time $t=0$. However, we must be cautious to conclude that state transitions tend to follow CPs in this case. Because of our requirement that the stateÕs posterior probability must exceed $80\%$ to detect an ensemble transition, the scored time of occurrence is likely to lag behind the correct time, which would skew the CPTA towards positive values. In any case, the CPTA peak is found around $t=0$ (Fig. \ref{Fig3}B) and shows that the largest proportion of state transitions co-occurs with a CP. Significance was established by comparison with a CPTA in a trial-shuffled dataset ($p<0.05$, t-test with Bonferroni correction; see Appendix~\ref{methods}). In the trial-shuffled dataset, the CPTA was flat indicating that the relation between CPs and ensemble transitions was lost (Fig. \ref{Fig3}B, blue trace). 
The fraction of neurons having CPs co-occurring with a state transition ranged from a single neuron to half of the ensemble (Fig. \ref{Fig3}C), indicating that more than one neuron, but not all neurons in the ensemble, co-activate during a state transition. In support of this result, we found that population firing rate modulations and single neurons modulations were partially correlated across transitions (Fig. \ref{Fig3}D, $\phi^2=0.21\pm0.04$, mean$\pm$SD). We explained the observed correlation by comparing it to three surrogate datasets (correlations in the four datasets were all significantly different; $\chi^2 (3)=110,\,p<10^{-10}$, Kruskal-Wallis one-way ANOVA with post-hoc multiple comparisons). The empirical correlation was significantly larger than in the case of a surrogate dataset in which all neurons changed their firing rates at random  ($\phi^2=0.01\pm0.01$, but significantly lower than in surrogate datasets with high levels of global coordination. A dataset with $90\%$ and $50\%$ of the neurons having co-occurring changes in firing rates yielded a $\phi^2=0.93\pm0.01$ and $\phi^2=0.66\pm0.02$, respectively (Fig. \ref{Fig3}D). Taken together, these results confirm that state transitions are due to a partial (and variable in size) co-activation of a fraction of neurons in the whole ensemble. 
Beyond unveiling the multi-state structure described above, inspection of the representative examples in Fig. \ref{Fig1} also provides indications on the spiking behavior of single neurons. Single neuronsÕ firing rates appear to be bi-stable in some cases and multi-stable in others. To quantify this phenomenon, we computed the firing rate distributions across states for each neuron. Fig. \ref{Fig3}E shows two representative examples featuring mixtures of distributions peaked at multiple characteristic firing rates (different states are color-coded). To find out the minimal number of firing rates across states for all recorded neurons, we conducted a conservative post-hoc comparison after a non-parametric one-way ANOVA of the firing rates for each neuron (see Appendix~\ref{methods}). The mean firing rates varied significantly across states for $90\%$ of neurons ($150/167$; Kruskal-Wallis one-way ANOVA, $p<0.05$), with at least $42\%$ ($70/167$) of all neurons being multi-stable, i.e., having three or more significantly different firing rate distributions across states (Fig. \ref{Fig3}F).
Altogether, these results demonstrate that ongoing activity in GC can be described as a sequence of multiple states in which a portion of neurons changes firing activity in a coordinated manner. In addition, our analyses show that almost half of the neurons switch between at least three different firing rates across states.   

\subsection{A spiking network model with a landscape of multi-stable attractors}

\begin{figure}[ht]
\begin{center}
\hspace*{-0.2cm}                                                           
\includegraphics[width=1.02\textwidth]{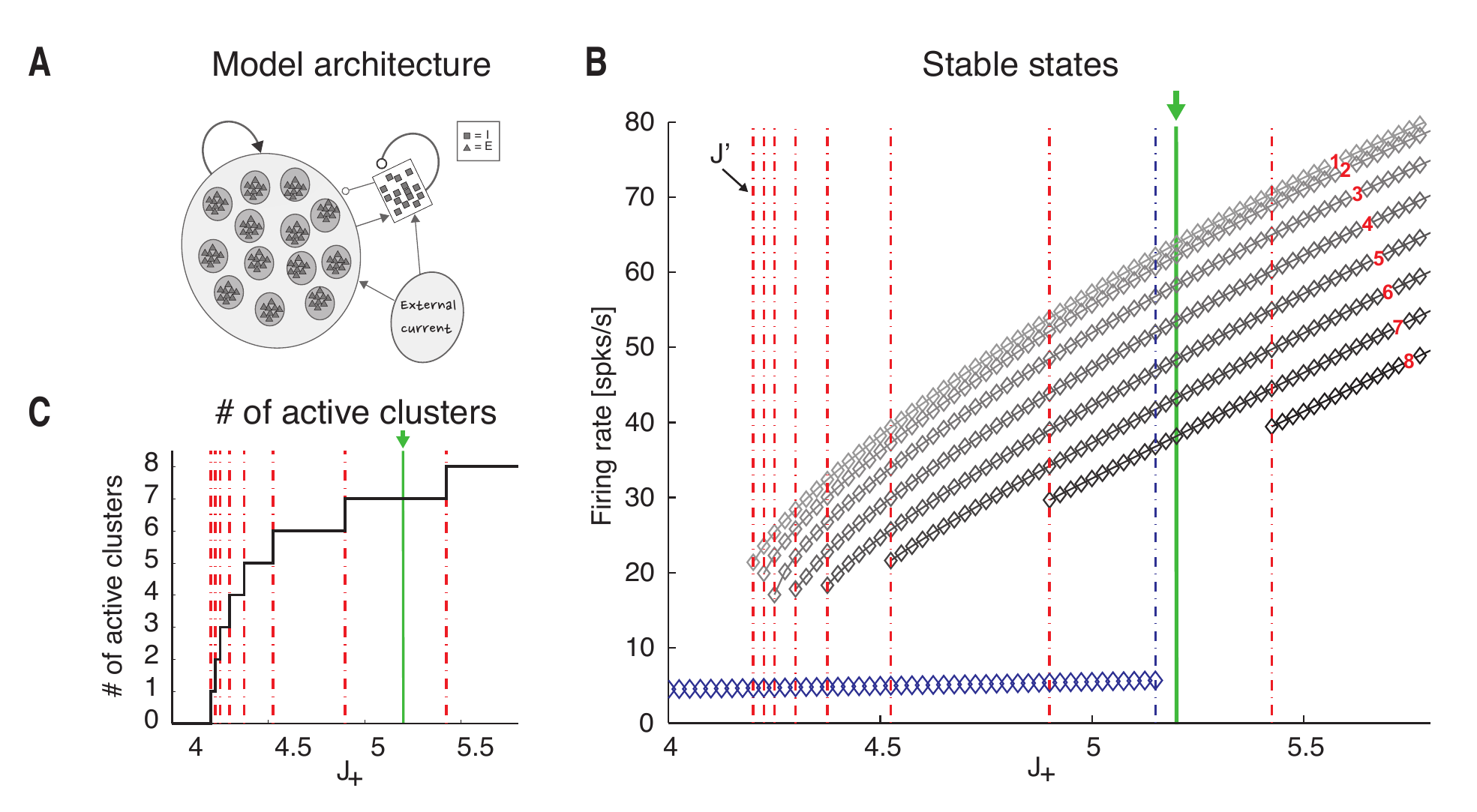}
\end{center}
\vspace*{-0.5cm}
\caption{{\it Attractor landscape of the spiking network model}. {\bf A}: network architecture. Excitatory (E, triangles) and inhibitory (I, squares) LIF neurons are recurrently connected. E neurons organized in $Q$ clusters; intra-cluster synaptic connections are potentiated (darker disks), inter-cluster connections are depressed (lighter disks). {\bf B}: mean field theory analysis of the network in panel {\bf A}. Firing rates for excitatory clusters attractor states (diamonds) are shown as a function of the intra-cluster potentiation parameter $J_+$ (in units of $J_{EE}$). Below the critical point $J'=4.2$ (i.e. left most vertical dotted line) the only stable state has low firing rate $\sim5$ spikes/s (blue diamonds). At $J_+=J'$, a bifurcation occurs whereby states with active clusters at higher firing rate coexist (upper diamond). There are in total $Q$ such states. As $J_+$ is further increased, states with $2$ or more active clusters exist. For each state, all active clusters have the same firing rate (reported on the vertical axis). The vertical green line at $J_+=5.2$ represents the value chosen for the spiking network simulations; vertical red lines represent critical points, where a new configuration appears with an increased number of active clusters. $X$-axis: intra-cluster synaptic potentiation ($J_+$); $Y$-axis: firing rate (spks/s).  {\bf C}: number of active clusters as a function of $J_+$ (notations as in panel {\bf B}). For $J_+=J'$ all attractor states have only one active cluster; for $J_+=5.2$ (green line) there are $7$ possible configurations of attractor states, with $1, 2, \ldots, 7$ active clusters, respectively. $X$-axis: intra-cluster synaptic potentiation ($J_+$); $Y$-axis: firing rate (spks/s).}
\label{Fig4}
\end{figure}

We have shown that ongoing activity in rat gustatory cortex is structured and characterized by sequences of metastable states. Biologically plausible models capable of generating such states, based on populations of spiking neurons, have been put forward \cite{DecoHugues2012,LitwinKumarDoiron2012,LitwinKumarDoiron2014}. However, the hallmark of these models is bi-stability in single neurons, which fire at either low or high firing rate, depending on the state of the network \cite{Amit1997b,Renart2007,Churchland2012}. Instead, a substantial fraction of GC neurons can exhibit $3$ or more stable firing rates (Fig. \ref{Fig3}F), raising a challenge to existing models. To address this issue, we developed a spiking network model capable of multi-stability, where each single neuron can attain several firing rates (i.e. $3$ or more). The network contains $4000$ excitatory (E) and $1000$ inhibitory (I) leaky integrate-and-fire (LIF) neurons (Fig. \ref{Fig4}A and Appendix~\ref{methods}) that are mutually and randomly connected. A fraction of E neurons form an unstructured (``background") population with mean synaptic strength $J_{EE}$. The majority of E neurons are organized into $Q$ recurrent clusters. Synaptic weights within each cluster are potentiated (with mean value of $\langle J\rangle=J_+ J_{EE}$, with $J_+>1$), whereas synapses between E neurons belonging to different clusters are depressed (mean value $\langle J\rangle=J_- J_{EE}$, with $J_-<1$). Inhibitory neurons are randomly interconnected to themselves and to the E neurons. All neurons also receive constant external input representing contributions from other brain regions and/or an applied stimulus (such as taste delivery to mimic our data). 
As explained in detail below, a theoretical analysis of this model using mean field theory predicts the presence of a vast landscape of attractors, and the existence of a multi-stable regime where each neuron can fire at several different firing rates. 

\subsection{Mean field analysis of the model}
A mean field theory analysis (see Appendix~\ref{methods}) predicts that, depending on the intra-cluster potentiation parameter $J_+$, the cortical network can exhibit configurations with one or more active clusters, where we use the term ``configuration" of the network to distinguish its admissible states of firing rate activities across neurons from the ensemble states revealed by an HMM analysis. A network configuration is fully specified by the number and identity of its active clusters, although we shall often focus on the number of active clusters regardless of their identity. The predicted population firing rates, as a function of $J_+$ and the number of active clusters in stable network configurations, are shown in Fig. \ref{Fig4}B. For very weak values of $J_+\gtrsim1$, the network has one stable configuration of activity (``background" configuration), where all E neurons fire at a low rate $\nu_L\sim5$ Hz (Fig. \ref{Fig4}B, blue diamonds left to the leftmost vertical red line at $J'=4.2$). As $J_+$ Increases beyond the first critical point $J'>1$, a bifurcation occurs (Fig. \ref{Fig4}B). At the bifurcation, a new set of attractors emerges characterized by a single cluster active at a higher firing rate $\nu_H$, while the rest of the network fires at $\approx\nu_L$ (gray diamonds between the two leftmost vertical red lines in Fig. \ref{Fig4}B). There are $Q$ possible such configurations, one for each cluster. As we increase $J_+$ in small steps, we cross several bifurcation points (vertical red lines in Fig. \ref{Fig4}B). To the right of each line, a new set of global configurations is added to the previous ones, characterized by an increasing number of simultaneously active clusters, whose firing rates depend on the number of active clusters. For example, for $J_+=5.2$ (green line), the network can be in one of $30$ configurations with only one active cluster at $\nu_H^{(1)}=64$ spikes/s; with $2$ active clusters at $ \nu_H^{(2) }=62$ spikes/s (with $\binom{30}{2}=435$ possible configurations); with $3$ active clusters at $\nu_H^{(3)}=58$ spikes/s (with $\binom{30}{3}=4060$ possible configurations), and so on. A configuration with $q$ out of $Q$ simultaneously active clusters can be realized in any one of $\binom{Q}{q}={Q!\over q!(Q-q)!}$ possible configurations, giving rise to a vast landscape of attractors. 
The firing rates $\nu_H^{(q)}(q\leq Q )$ depend on the number $q$ of simultaneously active clusters in decreasing fashion, i.e.  
$\nu_H^{(1)}>\nu_H^{(2)}>\ldots >\nu_H^{(q)}>\nu_L$.
The lower firing rate per cluster in a larger number of active clusters is the consequence of recurrent inhibition, since the whole network is kept in a state of global balance of excitation and inhibition \cite{van1996chaos,vreeswijk1998chaotic,Renart2007}. For each maximal number $q_\textrm{max}$ of allowed active clusters in the range of $J_+$ in Fig. \ref{Fig4}B, any subset of $q_\textrm{max}$ clusters could in principle be active, and the network can be in any one of the global configurations characterized by a number of active clusters compatible with the intra-cluster potentiation $J_+$. The number of different firing rates per neuron in the stable configurations (diamonds) are plotted as a function of $J_+$ in Fig. \ref{Fig4}C. 
It is important to realize that these configurations are stable, and would thus persist indefinitely, only in a network with an infinite number of neurons. Moreover, configurations with the same number of active clusters are equally likely (for a given value of $J_+$) only if all clusters contain exactly the same number of neurons. We show next in model simulations that, in a network with a finite number of neurons, the persistent configurations are spontaneously destabilized, resulting in the network going through a sequence of metastable states, as observed in the data. Moreover, slight variations in the number of neurons per cluster greatly decrease the number of observed configurations, also in keeping with the data. 

\subsection{Model simulations: asynchronous activity and multi-stable regime in the spiking network model }

\begin{figure}[b]
\begin{center}
\hspace*{-0.85cm}                                                           
\includegraphics[width=1.13\textwidth]{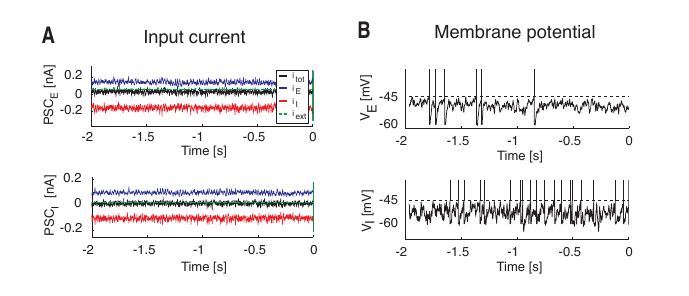}
\end{center}
\vspace*{-0.5cm}
\caption{{\it Dynamics of the spiking network model during ongoing activity}. Simulation of the network in Fig. \ref{Fig4} with $4000$ excitatory and $1000$ inhibitory LIF neurons, $Q=30$ clusters at intra-cluster synaptic potentiation $J_+=5.2$. {\bf A}: incoming PSC to an excitatory (PSCE, upper plot) and an inhibitory (PSCI, lower plot) neuron: EPSC (blue trace), IPSC  (red trace), external current (green line), and total current (black trace), are in a balanced regime. $X$-axis: time (s); $Y$-axis: post-synaptic current (nA). {\bf B}: membrane potential from representative excitatory ($V_E$, upper plot) and inhibitory ($V_I$, lower plot) neurons (vertical bars: spikes, horizontal dashed lines: threshold for spike emission). $X$-axis: time (s); $Y$-axis: membrane potential $V$ (mV). For illustration purposes, $V$ was linearly transformed to obtain the threshold for spike emission at $-45$ mV and the reset potential after a spike at $-60$ mV.}
\label{Fig5}
\end{figure}

\begin{figure}[ht]
\vspace{-1cm}                                                        
\begin{center}
\hspace*{-0.5cm}                                                           
\includegraphics[width=1.05\textwidth]{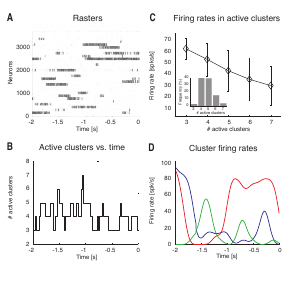}
\end{center}
\vspace*{-1.5cm}
\caption{{\it Dynamics of the spiking network model during ongoing activity}. {\bf A}: a representative rasterplot from excitatory clustered neurons (each dot represents a spike, background population not shown). Clusters of neurons that are currently active appear as darker regions of the raster. $X$-axis: time (s); $Y$-axis: neuron index. {\bf B}: time course of the number of active clusters from the representative trial in panel {A}. $X$-axis: time (s); $Y$-axis: number of active clusters.  {\bf C}: average firing rates in the active clusters as a function of the number of active clusters across all simulated sessions (bars: S.D.). $X$-axis: number of active clusters; $Y$-axis: average cluster firing rate (spks/s). Inset: occurrence of states with different counts of active clusters for $5\%$ stimulus amplitude. $X$-axis: number of active clusters; $Y$-axis: frequency of occurrence (\% of total time). {\bf D}: instantaneous cluster firing rate in three representative clusters (red, blue, and green lines) from trial in panel {A}. $X$-axis: time (s); $Y$-axis: firing rate (spks/s).}
\label{Fig6}
\end{figure}

We tested the predictions of the theory (Fig. \ref{Fig4}) in a simulation of the model in the regime corresponding to $J_+=5.2$ (green line in Fig. \ref{Fig4}B). Ninety percent of the E neurons were assigned to $Q=30$ clusters with potentiated intra-cluster connectivity $J_+ J_{EE}$. The clusters had a variable size with a mean number of $120$ neurons and $1\%$ standard deviation. In the ``background" configuration, the network was in the balanced regime as shown in Fig. \ref{Fig5}A for one E (top) and one I neuron (bottom), respectively. In this regime, the excitatory (blue), inhibitory (red), and external (green) post-synaptic currents sum up to a total current with mean close to zero (black trace) and large variance. This E/I balance results in irregular spike trains originating from substantial membrane potential fluctuations (Fig. \ref{Fig5}B). In line with this, the network had a global synchrony index of $\Sigma\approx0.01\sim1/\sqrt{N}$ \cite{Golomb2000}, which is the signature of asynchronous activity (see Appendix~\ref{methods}). The fluctuations generated in the network turned the stable configurations predicted by the mean field analysis (Fig. \ref{Fig4}B) into configurations of finite duration, and the network dynamically sampled different metastable configurations with different patterns of activations across clusters (Fig. \ref{Fig6}A), reminiscent of the sequences shown in Fig. \ref{Fig1}. In each configuration, $q$ out of $Q$ clusters were simultaneously active at firing rate $\nu_H^{(q)}$. In the example of Fig. \ref{Fig6}A, the number of active clusters ranged from $3$ to $7$ at any give time (Fig. \ref{Fig6}B). The number of active clusters across all sessions was approximately bell-shaped around $4.8\pm0.9$ (mean$\pm$SD; Fig. \ref{Fig6}C, inset); the configurations with appreciable frequency of occurrences had $3$ to $7$ active clusters, with firing rates ranging from a few to about $70$ spikes/s. Configurations with $1$ or $2$ active clusters, although predicted in mean field (Fig. \ref{Fig4}B), were not observed during simulations, presumably due to low probability of occurrence and/or inhomogeneity in cluster size (more on this point below). Moreover, the average population firing rate in each cluster was inversely proportional to the number of active clusters, as predicted by the theory (Fig. \ref{Fig6}C). The actual firing rates as a function of the number of active clusters were also in good agreement with the values predicted by mean field theory (Fig. \ref{Fig4}B). As the network hops across configurations, the firing rates of single neurons also change over time, jumping to values that depend on the number of co-active clusters at any given time (Fig. \ref{Fig6}D; different clusters are in different colors). These results suggest that this model can provide an explanation of the ongoing dynamics observed in the data, as we show next.

\subsection{Ensemble states in the network model}

\begin{figure}[ht]
\begin{center}
\vspace*{-1.5cm}                                                           
\hspace*{-0.8cm}                                                           
\includegraphics[width=1.1\textwidth]{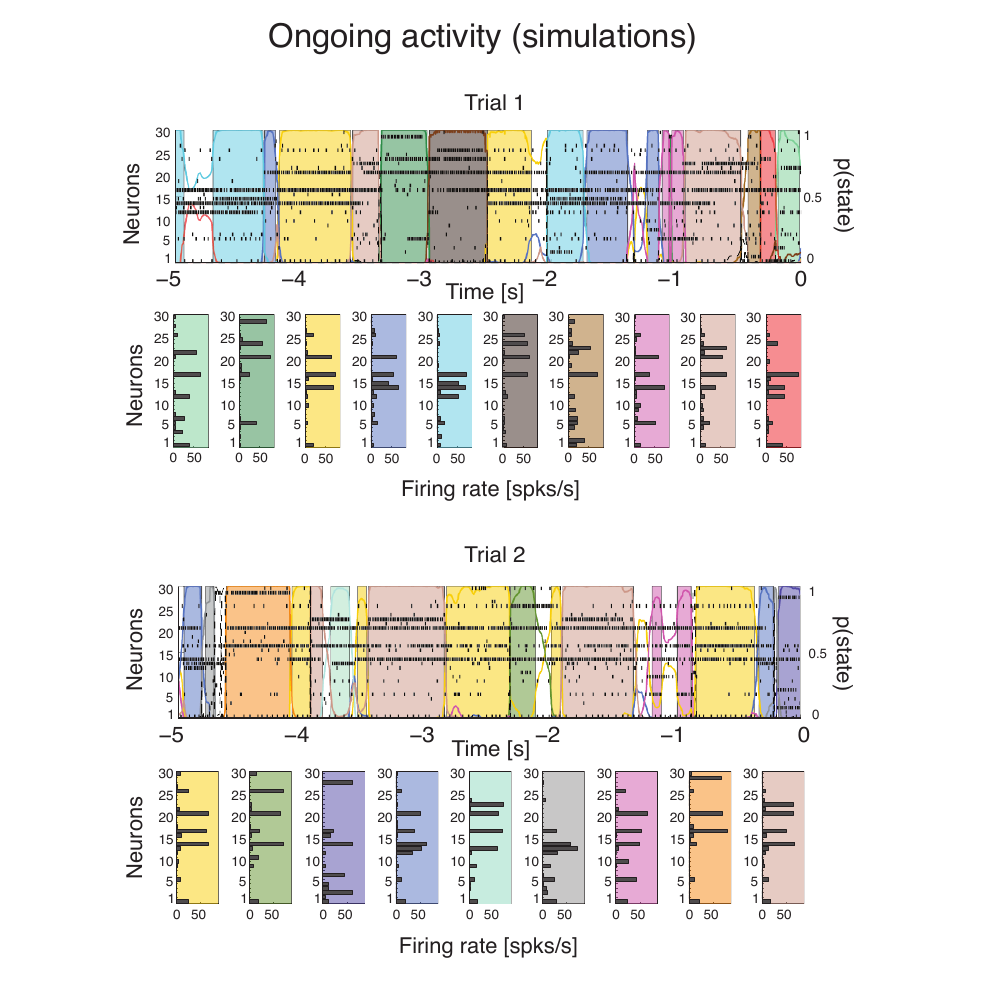}
\end{center}
\vspace*{-0.8cm}
\caption{{\it Ongoing activity in the model is characterized by sequences of states}. Two representative trials from an ensemble with 30 neurons from Fig. \ref{Fig6}A, one neuron per cluster (notations as in Fig. 1A). Raster plots (upper panel) are overlaid with HMM states, together with their firing rates across neurons (lower panel). Upper panel: X-axis, time preceding the next event (0 = stimulus delivery); Y-axis, left: ensemble neuron index; right: probability of HMM states. Lower panel: X-axis, firing rates (spks/s); Y-axis: ensemble neuron index. }
\label{Fig7}
\end{figure}

The model configurations with multiple active clusters could be the substrate for the states observed in the data. To show this, we performed an HMM analysis on $30$ simulated model sessions, each containing $40$ ongoing trials per session, each trial lasting $5$ seconds (Fig. \ref{Fig7}). Since all neurons in the same cluster tend to be active or inactive at the same time (Fig. \ref{Fig6}A), it is sufficient to consider one neuron per cluster in the HMM analysis. Fig. \ref{Fig7} shows examples of segmentation in states for a simulated ensemble of $30$ neurons (each randomly chosen from a different cluster). The duration and distribution of states in the sequences depended on a number of factors, among which the amount of heterogeneity in cluster size (larger clusters were more likely to be in an active state), the synaptic time constants and the intra-cluster potentiation parameter $J_+$. For our set of parameters we found $34\pm4$ states per session, ranging from $29$ to $39$. Examples of these states are shown in Fig. \ref{Fig7}. 
Note that different configurations with the same number, but different identity, of active clusters would produce different states, thus in principle our network could produce a very large number of states if the clusters would transition independently. Three factors limit this number to the actually observed one: 1) due to the networkÕs organization in clusters competing via recurrent inhibition, the number of predicted configurations that are stable have between $1$ and $7$ active clusters for $J_+=5.2$, as predicted by the theory (Fig. \ref{Fig4}B, diamonds on the green vertical line). No configurations with $8,\ldots, 30$ clusters are stable, because activation of one cluster will tend to suppress, via recurrent inhibition, the activation of other clusters; 2) not all predicted configurations have the same likelihood of being visited by the dynamics of the network; in our example, the probability of observing configurations with less than 3 active clusters in model simulations have very low probability (Fig. \ref{Fig6}C), further reducing the total number of visited configurations; 3) due to a slight heterogeneity in cluster size ($1\%$ variance across clusters), configurations with the larger clusters active tend to be sampled the most (i.e., they tend to recur more often and last longer), presumably due to a larger basin of attraction and/or a deeper potential well in the attractor landscape. All these factors working together produce an itinerant network dynamics among only a handful of configurations (compared with the large number possible a priori), that in turn originate only a handful of ensemble states in an HMM analysis of the same simulated data (Fig. \ref{Fig7}).
 The presence of recurrent inhibition and the heterogeneity in cluster size are also responsible for a higher degree of cluster co-activation than expected by chance had the clusters been independent. This was tested by comparison with a surrogate dataset obtained by random trial shuffling of the simulated data. The distributions of the number of cluster co-activations in 50 ms bins in the original vs. the shuffled data were significantly different (Kolmogorov-Smirnov test, $N_\textrm{eff}=3\times10^4,\,p<10^{-12}$ \cite{Press2007}). In particular, activations of single clusters were more frequent in the trial-shuffled dataset whereas co-activations of multiple clusters were more frequent in the original model simulations (one-way ANOVA with Bonferroni corrected multiple comparisons, $p<0.01$). Finally, the average number of different configurations was significantly larger in the trial-shuffled dataset (Wilcoxon rank-sum test across sessions, $z=-2.4,\,p<0.05$). Taken together, these results show that the model network, although producing asynchronous activity hence low pairwise spike train correlations (not shown), induces more coordinated clustersÕ co-activations and a much more limited number of network configurations than expected by chance. In turn, this is compatible with a limited number of states as revealed by a HMM analysis (Fig. \ref{Fig7}). 
 
\begin{figure}[t]
\begin{center}
\hspace*{-0.8cm}                                                           
\includegraphics[width=1.1\textwidth]{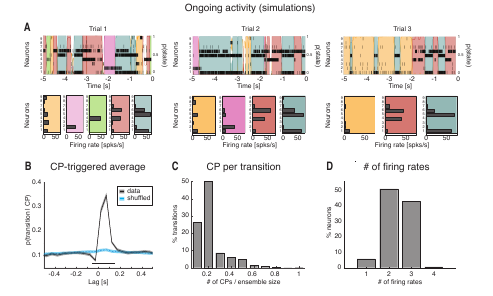}
\end{center}
\vspace*{-0.5cm}
\caption{{\it Ongoing activity in the model is characterized by sequences of states}. {\bf A}: three representative trials of simulated data after keeping only 9 randomly chosen neurons, together with a representation of the corresponding HMM states (lower panel); notations as in panel A. {\bf B}: CP-triggered average (CPTA) of ensemble transitions for the simulated dataset (black) and for a trial-shuffled dataset (blue); same conventions as in Fig. \ref{Fig3}B. Thick black line: $p<0.05$ (t-test with multiple-bin Bonferroni correction). X-axis: time lag from CP (s); Y-axis: probability of a transition given a CP at $t=0$. {\bf C}: fraction of transitions co-occuring with CPs within the $200$ ms significant window from panel B. X-axis: number of single neuronsÕ CPs divided by the ensemble size; Y-axis: fraction of transitions co-occurring with CPs (\%). {\bf D}: number of different firing rates per neuron across all sessions and neurons: $44\%$ of all model neurons had three or more different firing rates across ensemble states (compare with Fig. \ref{Fig3}F). X-axis: minimal number of different firing rates across states; Y-axis: fraction of neurons (\%).}
\label{Fig8}
\end{figure}

Note that a HMM analysis on $30$ simulated neurons taken from $30$ different clusters is bound to produce a larger number of states than detected in the data, because in the latter only ensembles of $3$ to $9$ simultaneously recorded neurons were available. Moreover, it is not guaranteed that the recorded neurons all came from different clusters. Thus, to facilitate comparison between model and data, we picked a distribution of ensemble sizes matching the empirical distribution ($3$ to $9$ neurons), and chose standard physiological values for all other parameters (Appendix~\ref{methods}). Three representative trials from an ensemble with $9$ neurons are shown in Fig. \ref{Fig8}A, revealing the re-occurrence of HMM states across trials, with $8.3\pm1.7$ states (mean$\pm$SD), ranging from $5$ to $12$ (a range much closer to the range of $3$ to $7$ states found in the data), and approximately exponential duration with mean of $239$ ms ($95\%$ CI: $[234, 243]$).  The number of states detected was correlated with the number of neurons in each state ($R^2=0.36,\,p<0.01$), as could be expected since the probability of detecting a state will increase in larger ensembles until neurons from all clusters are sampled. The model states matched other characteristic features of the data: state transitions followed CPs with a similar shape of the CPTA (Fig. \ref{Fig8}B; compare with Fig. \ref{Fig3}B) and were often congruent with the co-modulation of a fraction of the neurons (Fig. \ref{Fig8}C; compare with Fig. \ref{Fig3}C). Moreover, we found that $44\%$ of the neurons had $3$ or more firing rates (Fig. \ref{Fig8}D), essentially the same fraction ($42\%$) found in the data (see Fig. \ref{Fig3}F). No additional tuning of the parameters was required to obtain this match.

\subsection{Comparison between ongoing and stimulus-evoked activity: reduction of multi-stability}

\begin{figure}[ht]
\begin{center}
\hspace*{-0.3cm}   
\includegraphics[width=1.03\textwidth]{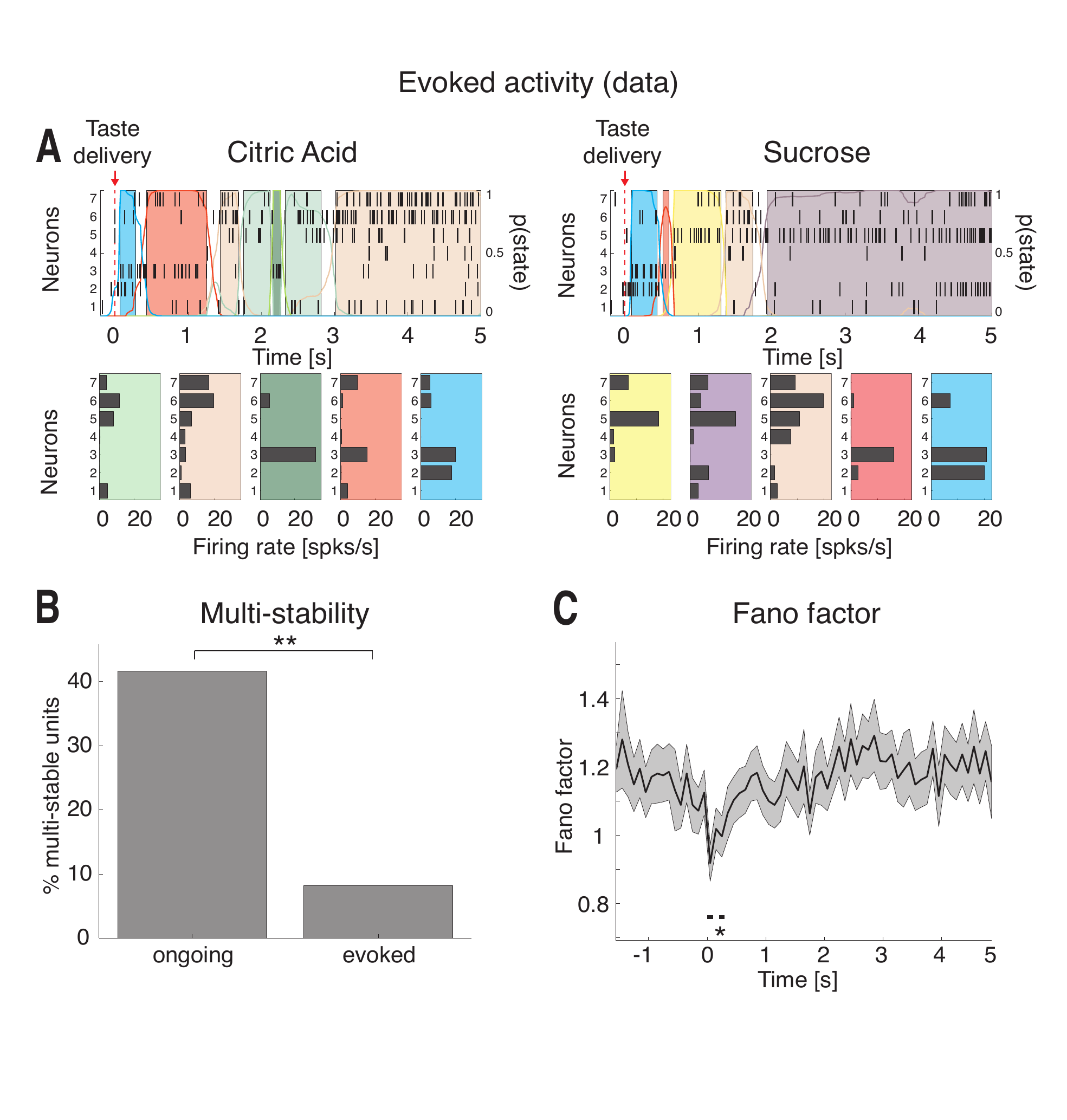}
\end{center}
\vspace*{-2cm}
\caption{{\it Taste-evoked activity: reduction of multi-stability and trial-by-trial variability in the experimental data}. {\bf A}: two representative examples of population rasters together with their segmentation in HMM states during evoked activity in GC, in response to citric acid (left) and sucrose (right) delivery, respectively (the red arrow marks taste delivery at $t=0$; remaining notations as in Fig. \ref{Fig1}A). Upper panel: $X$-axis, time relative to taste delivery ($0 =$ stimulus delivery); $Y$-axis, left, ensemble neuron index, right, posterior probability of HMM states. Lower panel: $X$-axis for lower panel: firing rates (spks/s); $Y$-axis, ensemble neuron index. {\bf B}: fraction of multi-stable neurons across all states during ongoing (left) and evoked activity (right) in GC ($**=p<0.01$, $\chi^2$ test). $X$-axis: ongoing and evoked conditions; Y-axis: fraction of multi-stable neurons (\%). {\bf C}: time course of the mean-matched Fano Factor (FF) in a time interval around taste delivery (occurring at time $t=0$) across all data. Shaded bounds: $95\%$ CIs. The thick horizontal black line indicates bins where the evoked FF is significantly different from baseline ($p<0.05$, see Appendix \ref{methods}). $X$-axis: time (s); $Y$-axis: Fano Factor.}
\label{Fig9}
\end{figure}

Previous work has proposed various relationships between patterns of ongoing activity and responses evoked by sensory stimuli, see e.g. \cite{Arieli1996,Kenet2003,FontaniniKatz2008,Luczak2009,Fiser2010,Abbott2011,Berkes2011}. We performed a series of analyses of the experimental and simulated data to determine whether the model network, developed to reproduce GC ongoing activity, captured the essential features of taste-evoked activity with no additional tuning of parameters. 
We first performed an HMM analysis on the data recorded during evoked activity in the $[0,5]$ s interval post-taste delivery, as in \cite{Jones2007}, see Fig. \ref{Fig9}A. We found a range of $4$ to $11$ states per taste across sessions (mean$\pm$SD: $7.2\pm1.6$) with an approximate exponential distribution of durations with mean 306 ms ($95\%$ CI: $[278, 334]$). However, during evoked activity only $8\%$ of the neurons had more than $2$ different firing rates across states compared to $42\%$ during ongoing activity (Fig. \ref{Fig9}B, $\chi^2 (1)=51,\,p<10^{-12}$). This suggests that a stimulus steers the firing rates of single neurons away from the multi-stable regime, hence reducing the range and number of different firing rates available. Moreover, in keeping with previous reports \cite{Churchland2010}, the stimulus caused a significant drop in trial-by-trial variability, as measured by the mean-matched Fano factor (see Appendix~\ref{methods} and Fig. \ref{Fig9}C).

\begin{figure}[ht]
\vspace{-0.7cm}                                                        
\begin{center}
\hspace*{-0.3cm}   
\includegraphics[width=1.03\textwidth]{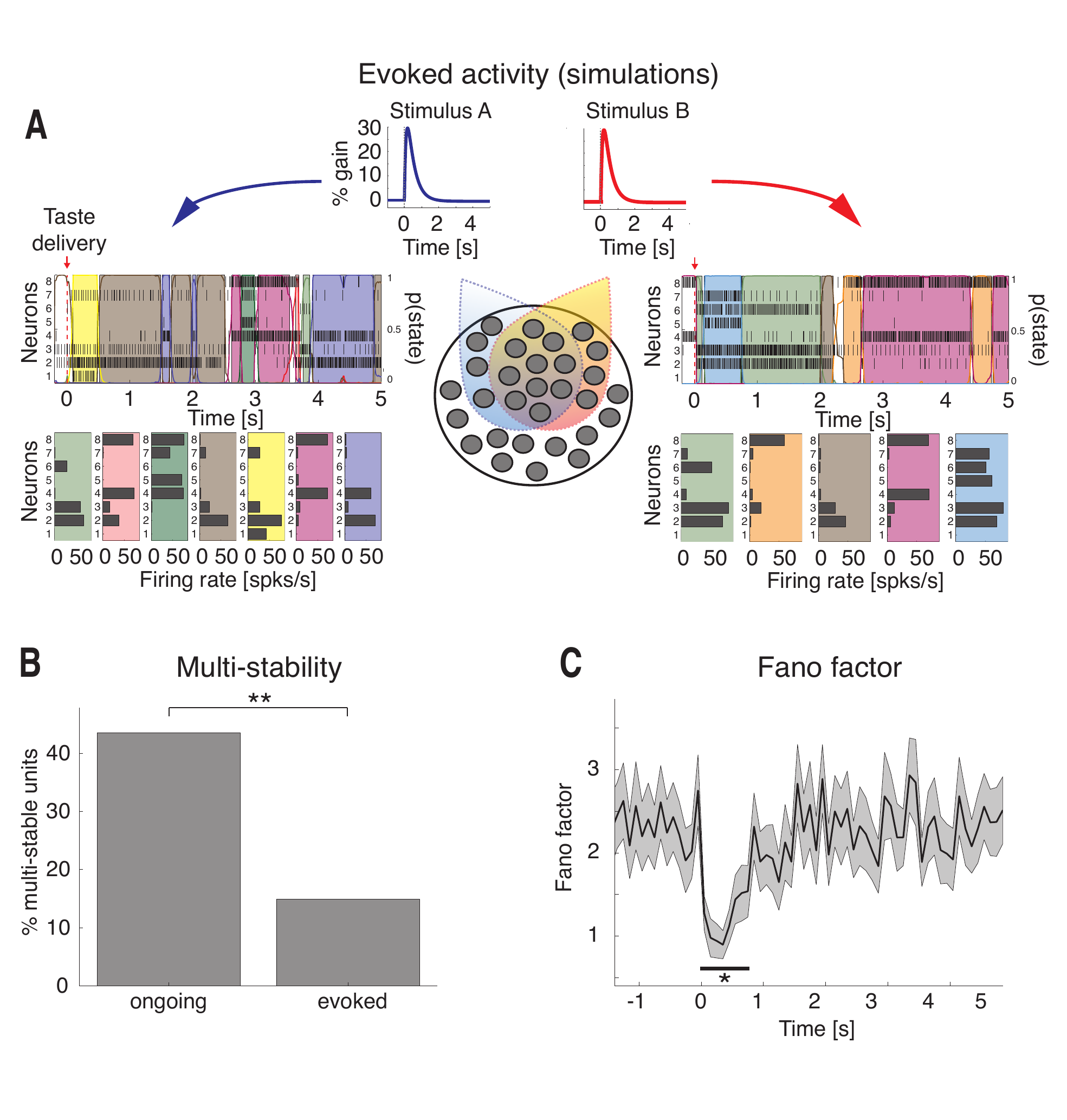}
\end{center}
\vspace*{-1cm}
\caption{{\it Taste-evoked activity: reduction of multi-stability and trial-by trial variability in the model}. {\bf A}: two examples of population rasters together with their segmentation in HMM states during evoked activity in the model network, in response to two surrogate stimuli. Clusters may be selective to more than one stimulus (central lower panel: dark circles and colored areas represent, respectively, individual clusters and their selectivity to the two different stimuli A (blue) and B (red)); remaining conventions as in Fig. \ref{Fig9}A. Model stimuli (central upper panel) consist of double exponentials peaking at $30\%$ of $\nu_{ext}$ ($50$ ms rise time and $500$ ms decay time). $X$-axis: time (s), $Y$-axis, stimulus amplitude (\%). {\bf B}: fraction of multi-stable neurons across all states during ongoing and evoked activity in the model ($**=p<0.01$). $X$-axis: ongoing and evoked conditions; $Y$-axis: fraction of multi-stable neurons (\%). {\bf C}: time course of the mean-matched Fano Factor in a time interval around stimulus presentation (occurring at time $t=0$) across all simulated sessions (same conventions as in Fig. \ref{Fig9}C). $X$-axis: time (s); $Y$-axis: Fano factor.}
\label{Fig10}
\end{figure}

To gain insight into the mechanism responsible for the observed reduction in multi-stability, we investigated whether this new configuration imposed by the stimulus is compatible with the attractor landscape predicted by the model (Fig. \ref{Fig4}B). We analyzed the behavior of the model network in the presence of a stimulus mimicking thalamic input following taste delivery (Appendix~\ref{methods}). Four stimuli were used (to mimic the $4$ tastants used in experiment), which differed only in the neurons they targeted, according to the same empirical distribution found in the data ($2$ representative stimuli are shown in the cartoon of Fig. \ref{Fig10}A). Specifically, the fractions of neurons responsive to $n=1,2,3$ or all $4$ stimuli were $17\%$ ($27/162$), $22\%$ ($36/162$), $26\%$ ($42/162$), and $35\%$ ($57/162$), respectively. All other model parameters were chosen as in the analysis of ongoing activity. 
With a biologically realistic stimulus peaking at $30\%$ of $\nu_\textrm{ext}$ (see Appendix~\ref{methods}), we found a range of $4$ to $17$ states per stimulus across sessions (mean$\pm$SD: $10.0\pm3.0$), with an approximate exponential distribution of state durations with mean $227$ ms ($95\%$ CI: $[219,235]$ ms).  Representative trials from simulated activity evoked by two of the four different stimuli are shown in Fig. \ref{Fig10}A, together with their segmentation in HMM states. As in the data, a significantly smaller fraction ($15\%$) of neurons had $3$ or more different firing rates across states compared to ongoing activity ($44\%$; Fig. \ref{Fig10}B, $\chi^2 (1)=24,\,p<10^{-5}$). This result was robust to changes in stimulus shape and amplitude; using different stimuli gave a comparable fraction of multi-stable neurons (either varying stimulus peak amplitude from $10\%$ to $30\%$ of $\nu_\textrm{ext}$, or using a box stimulus; not shown, see Appendix~\ref{methods}). The model also captured the stimulus-induced reduction in trial-by-trial variability found in the data (Fig. \ref{Fig10}C), as measured by the mean-matched Fano Factor, confirming previous results found with similar model networks \cite{DecoHugues2012,LitwinKumarDoiron2012}.

\begin{figure}[ht]
\vspace{-1cm}                                                        
\begin{center}
\hspace*{-0.3cm}   
\includegraphics[width=1.0\textwidth]{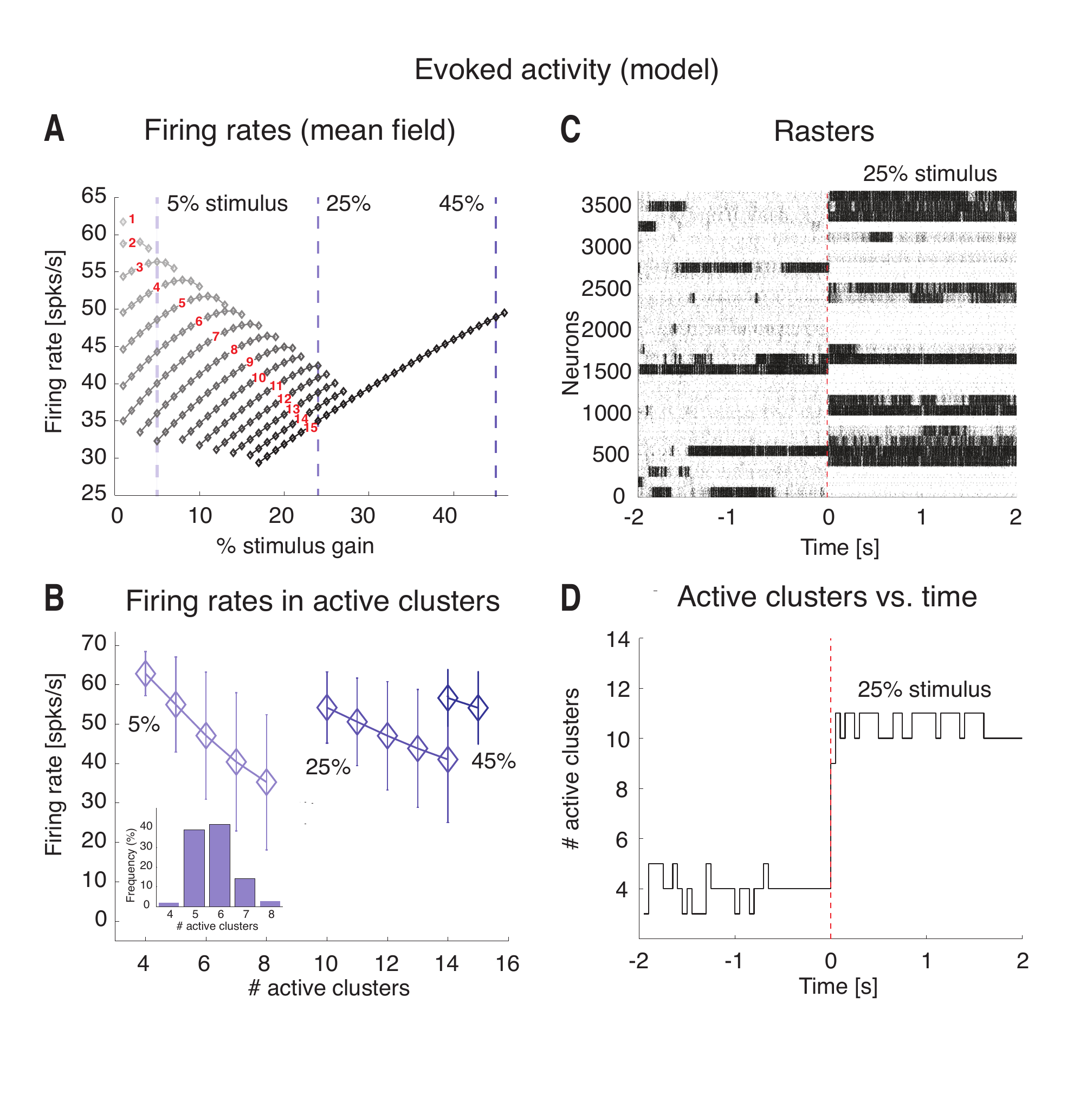}
\end{center}
\vspace*{-1.8cm}
\caption{{\it Mean field predictions of evoked activity and comparison with model simulations}. {\bf A}: attractor states of the model during evoked activity predicted by mean field theory for $J_+=5.2$ and as a function of stimulus amplitude. Diamonds: mean firing rate in active clusters for each attractor state, featuring the number of active clusters indicated by the red numbers (see also Fig. \ref{Fig4}B). Stimulus amplitude varied from $1\%$ to $45\%$ of $\nu_{ext}$. Vertical dashed lines: attractor states for stimuli of $5\%$, $25\%$ and $45\%$. The firing rate range decreases as the stimulus amplitude is increased, and it reduces to a single configuration with $15$ active clusters for stimuli $> 30\%$. $X$-axis: stimulus increase (\%); $Y$-axis: firing rate (spks/s). {\bf B}: mean firing rates in active clusters in simulations of the model for the three stimuli marked by vertical lines in panel A (vertical bars: S.D.). $X$-axis: number of active clusters; $Y$-axis: average cluster firing rate (spks/s). Inset: frequency of configurations with different numbers of active clusters for $5\%$ stimulus. $X$-axis: active clusters; $Y$-axis: frequency of occurrence (\% of total simulation time across all simulations). {\bf C}: rasterplot of spike trains from excitatory clustered neurons (background excitatory population not shown) in a representative trial encompassing both periods of ongoing and evoked activity (compare to Fig. \ref{Fig6}A), using a ÔboxÕ stimulus with amplitude at to $30\%$ of $\nu_{ext}$ in the $[0,2]$ s interval (ongoing activity corresponds to the $[-2,0]$ s interval). Vertical red line: stimulus onset. $X$-axis: time (s); $Y$-axis: neuron index. {\bf D}: number of active clusters vs. time (representative trial, panel C; same notation as in Fig. \ref{Fig6}B). $X$-axis: time (s); $Y$-axis: active clusters. }
\label{Fig11}
\end{figure}

Both the empirical data and the behavior of the network under stimulation could be explained with the same theoretical analysis used to predict ongoing activity (Fig. \ref{Fig4}). We repeated the mean field analysis for a fixed value of $J_+=5.2$ (green vertical line and arrow in Fig. \ref{Fig4}B) but varying the stimulus amplitude. The results are shown in Fig. \ref{Fig11}A. The number of active clusters tends to increase with stimulus amplitude, while the configurations with fewer active clusters gradually disappear. Moreover, one observes a reduction in the range of firing rates across states until, for a stimulus amplitude $\gtrsim30\%$ of $\nu_\textrm{ext}$, only configurations with $15$ active clusters are theoretically possible. Computer simulations confirmed these predictions (Fig. \ref{Fig11}B): although the match between predicted and simulated firing rates was not perfect (presumably due to finite size effects and the approximations used for synaptic dynamics), the predicted narrowing of the firing rate range for stronger stimuli was evident. As with ongoing activity, we also found that the network spent most of its time visiting configurations with an intermediate number of active clusters among all those theoretically possible predicted in mean field (inset of Fig. \ref{Fig11}B), suggesting, again, that different network configurations have different basins of attraction, with larger basins corresponding to longer durations. An example simulation for a stimulus amplitude of $30\%$ of $\nu_\textrm{ext}$ is shown in Fig. \ref{Fig11}C, together with the running score of the number of active clusters (Fig. \ref{Fig11}D).
In conclusion, stimulating a cortical network during a state of multi-stable ongoing activity reduces not only the trial-by-trial variability, as previously reported, but also multi-stability, defined as the single neuronsÕ ability to exhibit multiple ($3$ or more) firing rates across states. This may be linked to a reduction in complexity of evoked neural representations \cite{Rigotti2013}.

\section{Discussion}
\label{discussion}

Taste-evoked activity in GC has been characterized as a progression through a sequence of metastable states, where each state is a collection of firing rates across simultaneously recorded neurons \cite{Jones2007}. Here we demonstrate that metastable states can also be observed during ongoing activity and report several quantitative comparisons between ongoing and evoked activity. The most notable difference is that single neurons are multi-stable during ongoing activity (expressing $3$ or more different firing rates across states), whereas they are mostly bi-stable during evoked activity (expressing at most $2$ different firing rates across states). 
These results were reproduced in a biologically plausible, spiking network model of cortical activity amenable to theoretical analysis. The network had dense and sufficiently strong intra-cluster connections, which endowed it with a rich landscape of stable states with several different firing rates. To our knowledge, this rich repertoire of network attractors and its specific modification under stimulation has not been reported before. The stable states become metastable in a finite network, wherein endogenously generated fluctuations induce an itinerant dynamics among ensemble states. As observed in the data, this phenomenon did not require overt external stimulation. The model captured other essential features of the data, such as the distribution of single neuronsÕ firing rates across states and the partial co-activation of neurons associated to each ensemble transition. Moreover, without additional tuning, the model reproduced the reduction of multi-stability and trial-by-trial variability at stimulus onset. 
The reduction in trial-by-trial variability confirms previous results obtained across several brain areas \cite{Churchland2010}, which were also reproduced in spiking models similar to ours \cite{DecoHugues2012,LitwinKumarDoiron2012,LitwinKumarDoiron2014}. As for multi-stability, our theory offers a mechanistic explanation of its origin and its stimulus-induced reduction. The theory predicts the complete abolition of multi-stability above a critical value of the external stimulation (Fig. \ref{Fig11}A), when the network exhibits only one firing rate across all neurons in active clusters. In the more plausible range of mid-strength stimulations, the model shows a reduction in multi-stability and a contraction of the range of firing rates. 

\subsection{External and internal sources of metastability in GC}
Metastable activity can be induced by an external stimulus as recently shown by \cite{MillerKatz2010,Miller2013} in a spiking network model. Their model relies on feed-forward connections between recurrent cortical modules responsible for taste detection, taste identification, and decision to spit or swallow. Each stage corresponds to the activation of a particular module, and transitions between stages are driven by stochastic fluctuations in the network. While this model reproduces the dynamics of taste-evoked activity, it relies on an external stimulus to ignite the transitions, and thus would not explain metastable dynamics during ongoing activity. Our model provides a mechanistic explanation of intrinsically generated state sequences via a common mechanism during both ongoing and evoked activity.
The segmentation of ongoing activity in a sequence of metastable states suggests that the network hops among persistent activity configurations that lose stability by mechanisms including intrinsically generated noise \cite{Shpiro2009,DecoHugues2012,DecoJirsa2012,LitwinKumarDoiron2012,MattiaSanchezVives2012,LitwinKumarDoiron2014}, some process of adaptation \cite{Giugliano2004,Treves2005,Giugliano2008,Shpiro2009}, or winnerless competition \cite{Rabinovich2001}. In our model and in previous models similar to ours \cite{DecoHugues2012,LitwinKumarDoiron2012,LitwinKumarDoiron2014}, the mechanism is internally generated noise that destabilizes the stable attractors predicted by mean field theory. 
In a finite network, the stable states become metastable due to fluctuations of neural activity similar to those typically observed in vivo \cite{Shadlen1994,Rauch2003,LaCamera2006}. These fluctuations are a consequence of the network being in the balanced regime \cite{van1996chaos,vreeswijk1998chaotic,Renart2007}, which produces irregular and asynchronous spiking activity \cite{Brunel1999,Renart2010}. In the network containing a finite number of neurons, this fluctuating activity can destabilize the stable states and produce hopping behavior among metastable states \cite{DecoHugues2012,DecoJirsa2012,LitwinKumarDoiron2012,MattiaSanchezVives2012,LitwinKumarDoiron2014}. 

\subsection{Relationship between ongoing and evoked activity}
The nature of the relationship between ongoing and evoked states remains elusive. One proposal for the role of ongoing activity is that it serves as a repertoire of representations sampled by the neural circuit during evoked activity \cite{Arieli1996,Kenet2003,Luczak2009}. According to this proposal, evoked states occupy a subset of ongoing states in a reduced representational space, such as the space of principal components, or the space obtained after multidimensional scaling \cite{Luczak2009}. This applies especially to the activity of auditory or somatosensory neurons during high and low activity states called ``UP" and ``DOWN" states \cite{MacLean2005,Bartho2009,Luczak2009,Luczak2013}. 
An alternative proposal regards ongoing activity as a Bayesian prior that incrementally adjusts to the statistics of external stimuli during early development \cite{Fiser2004,Berkes2011}. In this framework, ongoing activity starts out as initially different from evoked activity, and progressively shifts towards it. 
Our modelÕs explanation of the genesis and structure of ensemble states has two main implications: if, on the one hand, ongoing and evoked activities are undoubtedly linked (they emerge from the same network structure and organization); on the other hand they may differ in important ways.  Given the networkÕs structure, only a handful of states are allowed to emerge through network dynamics, which might suggest that ongoing activity constrains the repertoire of sensory responses. However, the same networkÕs structure, especially its synaptic organization in potentiated clusters, can be obtained as the result of learning external stimuli \cite{Amit1997b,Amit2003,Tkacik2010,LitwinKumarDoiron2014}, so from this point of view it is the realm of evoked responses that defines the structure of ongoing activity. In any case, even though originating from the same network structure and mechanisms, evoked and ongoing activity can be characteristically different. They may differ in the number of ensemble states or firing rate distributions across neurons, resulting in only a partial overlap when analyzed as in \cite{Luczak2009} (not shown). Stimuli drive neurons at higher and lower firing rates than typically observed in ongoing activity, preventing ongoing activity from completely encompassing the evoked states. This is a direct consequence of our finding that activity in GC is not compatible with UP and DOWN states, both in terms of the dynamics of state transitions (Fig. \ref{Fig3}D) and in terms of the wider range of firing rates attained by single neurons (multi-stability, Fig. \ref{Fig3}F).
The observed differences between ongoing and evoked activity can be understood thanks to our spiking network model. This model shows how the presence of a stimulus induces a change in the landscape of metastable configurations (Fig. \ref{Fig11}A). This change depends on the strength of the stimulus and is incremental, quenching the range and the number of the active clusters in admissible configurations. Beyond a critical point ($\gtrsim30\%$ stimulus strength in Fig. \ref{Fig11}A), the only states left are in configurations where all stimulated clusters are simultaneously active, and evoked states are intrinsically different from ongoing states. Although they still relate to ongoing states due to their common network origin, evoked states contain information that is not available in the ongoing activity itself (and indeed, this information can be used to decode the taste, see e.g. \cite{Jones2007,samuelsen2012effects, jezzini2013processing}). 

\subsection{Mechanisms of itinerant multi-stable activity}
We have shown that our network model, although balanced, produces coordinated co-activation across clusters, mostly because of recurrent inhibition and heterogeneity in cluster size. Recurrent inhibition reduces the number of clusters that can be simultaneously active, whereas heterogeneity in cluster size inflates the probability of visiting larger clusters at the expense of smaller ones, thus initiating recurrent sequences of states. In turn, this results in more coordinated cluster dynamics and a much more limited number of network configurations than expected by chance. 
It is tempting to infer, on the basis of the results presented in Figures \ref{Fig3} and \ref{Fig8}, a similar degree of partial coordination in the experimental data during ongoing activity. Given the limited size of our ensembles and the limited number of neurons participating in state transitions, it is difficult to determine whether the experimental data reflect the same degree of coordination predicted by the model. While recent evidence of clustered subpopulations in the frontal and motor cortex of behaving monkeys seem to support our model \cite{kiani2015natural}, future experiments making use of high-density recordings may provide more accurate measurements of the degree of coordination among clusters. It is worth noting that the latter may not be fixed but rather depend on the behavioral state of the animal. During sleep or anesthesia, a spiking network may experience more synchronous global states, while during alertness it may exhibit intermediate degrees of coordination such as those predicted by our model. Similarly, evoked activity might display a more global degree of coordination than ongoing activity. Depending on the strength of inter-cluster connections and the size of individual clusters, our model can produce networks with different degrees of coordination and thus provide a link between behavioral states and mechanisms of coordination. 
Finally, our model offers a novel mechanism for multi-stable firing rates across states, i.e., the observation that approximately half of our neurons have $3$ or more firing rates across states. Even though a network with heterogeneous synaptic weights could have a distribution of different firing rates across different neurons \cite{Amit1997a}, each neuron would only be firing at two different firing rates (bi-stability). This is at odds with multi-stability. In our model, having several firing rate patterns in each single neuron implies that the network visits states with different numbers of active clusters, where the firing rate of each cluster depends on the number of active clusters at any given time. 

State sequences seem to encode gustatory information and are believed to play a role in taste processing and taste-guided decisions \cite{Jones2007,MoranKatz2014,Mukherjee2014}. Potential mechanisms to read out such sequences in populations of spiking neurons are being investigated \cite{Kappel2014}. In addition to taste coding, multi-stable activity across a variety of metastable states may enrich the network's ability to encode high dimensional and temporally dynamic sensory experiences. The investigation of this possibility and its link to the dimensionality of neural representations \cite{Rigotti2013} is left for future work.

\section*{Acknowledgments}

This work was supported by a National Institute of Deafness and Other Communication Disorders Grant K25-DC013557 (LM), by the Swartz Foundation Award 66438 (LM), by National Institute of Deafness and Other Communication Disorders Grant R01-DC010389 (AF), by a Klingenstein Foundation Fellowship (AF), and partly by a National Science Foundation Grant IIS1161852 (GLC). We thank Dr. Stefano Fusi and the members of the Fontanini and La Camera laboratories for useful discussions.

\bibliography{bib}
\bibliographystyle{JHEP}

\end{document}